\documentclass[12pt,reqno]{amsart}
\usepackage{graphics,graphicx,amsmath,amssymb,epsfig,stmaryrd,color,flafter,xspace,lscape,threeparttable,pdfpages}
\usepackage{csquotes,amsfonts,amsthm,geometry,array,booktabs,latexsym,caption,subfig,multirow,tabularx,rotating,bigstrut}
\usepackage[font=small,labelfont=bf]{caption}
\DeclareCaptionFont{tiny}{\tiny}
\everymath{\displaystyle}

\RequirePackage{setspace,indentfirst,epsfig,psfrag,ifthen,ifpdf}
\usepackage{changepage}

\newtheorem{theorem}{Theorem}
\theoremstyle{plain}

\newtheorem{assumption}{Assumption}

\newtheorem{corollary}{Corollary}

\newtheorem{definition}{Definition}
\newtheorem*{definition*}{Definition}

\newtheorem{proposition}{Proposition}

\numberwithin{equation}{section}

\begin{document}

\title[Sharp bounds for the Roy model]{Sharp bounds and testability of a Roy model \\ of STEM major choices}
\author{Isma\^^22el Mourifi\'e}\address{University of Toronto}
\author{Marc Henry}\address{The Pennsylvania State University} 
\author{Romuald M\'eango}\address{Munich Center for the Economics of Ageing at the Max-Planck Insitute for Social Law and Social Policy}
\noindent \date{\scriptsize{The first version is of 22 April 2012. The present version is of \today. This research was supported by SSHRC Grants 410-2010-242, 435-2013-0292 and 435-2018-1273, NSERC Grant 356491-2013, and Leibniz Association Grant SAW-2012-ifo-3. The research was conducted in part, while Marc Henry was visiting the University of Tokyo and Isma\^^22el Mourifi\'e was visiting Penn State and the University of Chicago. The authors thank their respective hosts for their hospitality and support. They also thank
D\'esir\'e K\'edagni, Lixiong Li, Karim N'Chare, Idrissa Ouili and particularly Thomas Russell and Sara Hossain for excellent research assistance. Helpful discussions with Laurent Davezies, James Heckman, Hidehiko Ichimura, Koen Jochmans, Essie Maasoumi, Chuck Manski, Ulrich M\^^22uller, Aureo de Paula, Azeem Shaikh
and very helpful and detailed comments from five anonymous referees, from numerous seminar audiences 
and the 2018 Canadian senate open caucus on women and girls in STEM are also gratefully acknowledged.
Correspondence address: Department of Economics, Max Gluskin House, University of Toronto, 150 St. George St., Toronto, Ontario M5S 3G7, Canada}}

{\scriptsize \begin{abstract}
We analyze the empirical content of the Roy model, stripped down to its essential features, namely sector specific unobserved heterogeneity and self-selection on the basis of potential outcomes. We characterize sharp bounds on the joint distribution of potential outcomes and testable implications of the Roy self-selection model under an instrumental constraint on the joint distribution of potential outcomes we call {\em stochastically monotone instrumental variable (SMIV)}. We show that testing the Roy model selection is equivalent to testing stochastic monotonicity of observed outcomes relative to the instrument. We apply our sharp bounds to the derivation of a measure of departure from Roy self-selection to identify values of observable characteristics that induce the most costly misallocation of talent and sector and are therefore prime targets for intervention. Special emphasis is put on the case of binary outcomes, which has received little attention in the literature to date. For richer sets of outcomes, we emphasize the distinction between pointwise sharp bounds and functional sharp bounds, and its importance, when constructing sharp bounds on functional features, such as inequality measures.
We analyze a Roy model of college major choice in Canada and Germany within this framework, and we take a new look at the under-representation of women in~STEM.
\end{abstract}
}

\maketitle

{\scriptsize \textbf{Keywords}: Roy model, partial identification, stochastic monotonicity, functional sharp bounds, inequality, college major, gender profiling, STEM, SMIV.

\textbf{JEL subject classification}: C31, C34, C35, I21, J24
}

{\footnotesize
\newpage




\section*{Introduction}

In a seminal contribution that is now part of the folklore of economics,
\cite{Roy:51} proposed a model of earnings with sorting on sector specific skills. Roy's objective was to provide a channel by which skills translate into earnings and to capture the idea that favorable sorting reduces earnings inequality. 
The simplicity of this mechanism and the richness of its implications turned the Roy model into one of the most successful tools in the analysis of environments, where skills and choices interact: they include the \cite{Gronau:74}-\cite{Heckman:74} labor supply model, the unionization model of \cite{Lee:78}, the model of education self-selection proposed by
\cite{WR:79}, sector selection in \cite{HS:85}, and the \cite{Borjas:87} immigration model. More recently, \cite{Lemieux:98} revisited the issue of inequality in the unionization model, \cite{MR:2004} used the Roy model to shed light on the recent evolution of the gender gap, \cite{CS:2007} to analyze the choice of surgical procedures and \cite{EHV:2014} to analyze benefits and costs of educational choices. The list is, of course, far from complete, but quite sufficient to show the enormous success of the Roy model.

In the original model, skills are jointly log normal and \cite{BG:78} show that under this assumption, the joint distribution of skills and the marginal distributions of potential earnings are identified. \cite{HH:90} further show that self-selection does indeed reduce aggregate inequality when skills are log normal and within sector inequality when skills have a log concave distribution.
Naturally, the effect of self-selection on outcome inequality remains empirically relevant when skills do not have a log concave distribution. However, the analysis of the nonparametric version of the Roy model, stripped down to the self-selection mechanism, has long been hampered by (lack of) identification issues. The \cite{Cox:72} and \cite{Tsiatis:75} comments on non-identifiability of competing risks imply that any continuous outcome distribution could be rationalized with independent sector-specific skills, so that the \cite{WR:79} notion of skill hierarchies loses empirical content.

One way to resolve this lack of identification issue, pioneered by \cite{HH:89}, \cite{HH:90}, is to bring in additional information to achieve identification, such as repeated cross sections, panel data, factor structure, exclusion restrictions and large support assumptions within restricted specifications of the model. A vast literature, both theoretical and empirical, followed this lead (see for instance \cite{Heckman:2001}, \cite{HV:2007}, \cite{HT:2008} or \cite{FT:2011} for recent accounts). In particular, recent developments in nonparametric inference in Roy and competing risks models can be found in \cite{KT:2007}, \cite{LL:2013}, \cite{BKT:2011}, and \cite{HM:2013}.
Another way to approach the issue, which was pioneered by \cite{Peterson:76} and which we follow here, is to recognize that, despite the identification failure because of self-selection, the Roy model is not devoid of empirical content.  

The object of the present article is to characterize this empirical content, with special emphasis on the joint distribution of potential outcomes and testability of the Roy selection mechanism. This implies considering distributional features of outcomes, which are important if one is to evaluate the effect of self-selection on wage inequality, as Roy initially intended; see \cite{CHN:2006} for a discussion. It further implies considering joint distributional aspects. As \cite{Heckman:92} noted, information on the joint distribution is necessary to evaluate welfare implications of policy changes that affect the relative price of skills in both sectors. Correlation between outcomes can be important to policy evaluation, as discussed in \cite{CHH:2002}, as can the difference between potential outcomes or the distribution of outcomes conditional on the chosen sector. In all such cases, the joint distribution of potential outcomes is the relevant object to characterize. We refer to \cite{Heckman:2010}, \cite{HSC:97}, and \cite{AH:2007} for in-depth discussions of this issue.

We devote a considerable amount of attention to the case with binary outcomes, which we call the {\em binary outcomes Roy model}. The reason is twofold. First, the identification failure is starker with binary outcomes, and the characterization of the joint distribution is easier to derive and explain in the binary case, before it is extended to the more general cases of discrete, continuous or mixed discrete and continuous outcomes. Second, the case of discrete outcomes has received very little attention in the Roy model literature, \cite{Poirier:80} and \cite{AHV:2005} being notable exceptions. Most of this literature
concerns the case of continuous outcomes and many applications,
where outcomes are discrete, fall outside its scope. They include
analysis of the effects of different training programs on the
ability to secure employment, of competing medical treatments
or surgical procedures on survival, and of competing policies
on schooling decisions in developing countries among numerous
others. The Roy model is still highly relevant to those
applications, but very little is known of its empirical content in
such cases.

We derive sharp bounds for the joint distribution of $(Y_0,Y_1)$, using techniques from \cite{GH:2011} and \cite{BMM:2011} -see also \cite{BMM:2012} and \cite{CRS:2011}.\footnote{\scriptsize \cite{HSC:97}, and later \cite{FR:2008}, \cite{FP:2010} and \cite{FW:2010} look at more general treatment effects models from a partial identification point of view and use rearrangement inequalities to derive bounds on the distribution of treatments effects, a feature of the joint distribution of potential outcomes, under conditions, where the marginal distributions of potential outcomes are identified.} Bounds do not cross, and the model is not testable, unless we observe variables that have a restricted effect on potential outcomes. A special case is that of selection shifters that are statistically independent of potential outcomes. Such variables have two major drawbacks in this framework. First, they are very elusive in important areas of application of this methodology. To take one classical example in the literature on returns to education, parental education, measures of school quality, and fees may be correlated with unobserved cognitive and non cognitive parental investments and is therefore unlikely to be independent of potential outcomes. Second, within the Roy model, given the sector selection mechanism, a variable that is independent of potential outcomes can only affect sector selection if potential outcomes are equal in both sectors, which severely restricts the extent of resulting variation in sector selection. 

To resolve both of these issues, we introduce {\em stochastically monotone instrumental variables}. They are selection shifters that are restricted to affect potential outcomes monotonically. For instance, parental education may not be independent of potential wages because of unobserved parental cognitive and non cognitive investments. However, it is unlikely that such additional investments will negatively affect potential future wages. Moreover, allowing for monotonic effects on potential outcomes resolves the second issue, since stochastically monotone instrumental variables may shift selection even when potential outcomes are different. Our stochastically monotone instrumental variable assumption is stronger than the \cite{MP:2000} {\em monotone instrumental variable} assumption, which only requires mean potential outcomes to be monotonic in the instrument, rather than the whole distribution. It is different from Constraint~(9) in 
\cite{BGIM:2007}, which is a restriction on the univariate distribution of realized wages, as opposed to the joint distribution of potential outcomes. This difference is crucial, when deriving bounds on joint distributional features in a sector selection model.

We derive the identified set for the joint distribution of potential outcomes in the binary outcomes Roy model under this assumption of stochastically monotone instrumental variable (hereafter {\em SMIV}), and we show that stochastic monotonicity of observed outcomes in the instrument summarizes all observable implications of the model. Hence a test of Roy selection behavior boils down to a test of stochastic monotonicity,  and can therefore be conducted with existing inference methods, as in \cite{LLW:2009}, \cite{DE:2012} and \cite{HLS:2016}. This provides a fully nonparametric alternative to tests of the Roy model proposed since \cite{HS:90} (see also \cite{Dahl:2002}, with multiple treatments and repeated cross-sections). Intuition about the relation between stochastic monotonicity and Roy selection can be gained from the following hypothetical scenario. Suppose two identical young women have higher economic prospects in non STEM fields than in STEM fields. One of them is induced to choose a STEM degree by a larger proportion of women on the STEM faculty in her region, whereas the other, who lives in a region with lower proportion of women on the STEM faculty, chooses a non-STEM field to maximize economic prospects. It will then appear that an increase in the proportion of women on the STEM faculty produces a decrease in observed outcomes, hence a rejection of monotonicity.

To alleviate the concern that rejection of Roy selection behavior may in fact be down to a rejection of the assumption that individuals are perfectly informed of their potential future outcomes at the time of sector selection, we derive sharp bounds for the joint distribution of potential outcomes for an imperfect foresight binary outcomes Roy model. In the latter, agents select the sector that maximizes the expectation of their outcome with respect to their information set at the time of decision. Since the model is rejected if and only if this identified set is empty, it allows us to summarize all observable implications of that version of the model as well. We also provide bounds for a measure of departure from Roy selection, which is constructed from the difference between the maximum potential outcome and the realized outcome (both being equal under Roy selection), and which, again, requires bounding the joint distribution of potential outcomes, as we do here, rather than marginal average outcomes, as is customary in the literature. These measures of departure from the Roy selection model serve to identify values of observable characteristics that induce the most costly misallocation of talent and field of study and are therefore prime targets for intervention.

We extend the test of Roy self-sorting and the measure of departure from Roy to more general discrete and continuous outcomes. When extending the analysis of the bounds on potential outcomes, distributional issues come to the fore. The classic \cite{Peterson:76} bounds are sharp for
$\mathbb P(Y_0\leq y)$ and $\mathbb P(Y_1\leq y)$, for each quantile $y$, but, as noted by \cite{Crowder:91}, they do not incorporate monotonicity and right-continuity restrictions on distribution functions. Hence they entail loss of information, when the object of interest involves densities, such as hazard rates, or functionals of the distribution, such as inequality measures. 
We provide a general characterization of the joint distribution and bounds on the marginal distributions of potential outcomes that are {\em functionally sharp}, in the sense that they incorporate slope restrictions. In this, we follow \cite{BM:97}, although the model specification, hence the bounds, are different. As in \cite{V_AMvS:2003}, \cite{BGIM:2007} and \cite{Stoye:2010} for different models, we apply the latter bounds to derive sharp bounds on inequality measures, which we show are more informative than would have been obtained from {\em pointwise sharp} bounds on the distributions of potential outcomes, such as Peterson bounds.

In the tradition of \cite{WR:79}, \cite{KLMT:79} and, more recently, \cite{EHV:2014}, we analyze returns to education through the lens of the Roy selection model. It is well documented, since at least \cite{JACT:89}, that major choice is an important determinant of labor market outcomes. The account in \cite{AAM:2016} shows that the literature on the determinants of and the returns to major choice is now substantial, including most notably \cite{Arcidiacono:2004}, \cite{beffy2012} and \cite{KLM:2015}. The STEM versus non-STEM classification has come to dominate the debate.
We therefore analyze a Roy model of choice of field of study, between STEM and non-STEM degrees, based on data from nationally representative surveys of university graduates in Canada and Germany. Following the recent literature on the subject, surveyed in \cite{KG:2017}, we focus on mathematics intensive fields, including economics, but excluding life sciences. 

We consider a Roy model of major choice, where the target labor market outcomes are wage, obtaining a permanent job by the year of the survey, or holding a job related to the field of study, respectively. Our main objective is to shed some insight onto the under-representation of women in STEM education and even more so in STEM jobs, the gender gap in STEM labor market outcomes, and the contribution of the STEM economy to rising wage inequality. If choices conform to the Roy self-sorting mechanism, only policies directed at ex-post wage discrimination are  likely to be effective in reducing inefficiencies, not policies directed at reducing gender profiling in major choice. Hence, in our investigation of the determinants of under-representation of women in STEM fields, we give prominence to testing Roy selection behavior. As the latter is not testable without covariate restrictions, we consider variation induced in major choices by parental education. There are reasons to doubt the validity of parental education according to Assumption~\ref{ass:IV} below, which requires independence of the instrument and the vector of potential outcomes. It is more reasonable to assume parental education level has a monotonic effect on potential outcomes, as prescribed by Assumption~\ref{ass:SMIV} below, which requires stochastic monotonicity of the distribution of potential outcomes, conditionally on the instrument. When testing whether women graduate choices conform to the Roy self-sorting mechanism, we also use the fraction of women in the faculty of STEM programs in the region and at the time of major choice as a SMIV instrument, based on the assumption that role models may not negatively affect future prospects for women graduates. We also perform our test of Roy self-sorting based on additional vectors of instruments incorporating local labor market conditions at the time of choice for robustness purposes, and find very little variation in test results.

Our tests of the Roy model for different choices of instruments and different employment related outcome variables reveal significant gender, racial and regional differences. 
We find a pattern of rejections of Roy self-sorting based on outcomes for white women in the former Federal Republic of Germany and the rest of Canada, and a lack of rejections for visible minorities and for white women from Qu\'ebec, white men from all of Canada and the former German Democratic Republic. Confidence intervals for measures of departure from Roy behavior reveal that in the case of white women from the former Federal Republic, for instance, rejection of Roy behavior seems to be driven by lower income women with high school educated mothers and middle income women with postgraduate educated mothers. Among groups, where Roy self-sorting is not rejected, comparisons of interquartile ranges for observed and counterfactual income distributions are inconclusive except in the case of white women in the former Democratic Republic and white men in Qu\'ebec, where self-sorting is found not to increase inequality, complementing the result for log-concave talent distributions in \cite{HH:90}.

The pattern of rejections of Roy self-sorting in major choice points to non labor market related determinants of choice. For instance, our results are consistent with a story involving gender profiling pushing white men in the West of Germany into STEM fields and white women in the West of Germany and in Canada out of STEM fields. They are also consistent with gender profiling being less prevalent in the former communist Germany. However, the results are also consistent with a story involving non pecuniary field preferences driving major choices of more privileged groups in more affluent regions, but not the choices of the more financially constrained.

\subsection*{Outline} The remainder of the paper is organized as follows. Section~\ref{sec:frame} details the general frame of analysis. Section~\ref{sec:binary} concerns sharp bounds and testability of the binary outcomes Roy model, and its version with imperfect foresight, under the assumption of stochastic monotonicity of potential outcomes relative to an instrument. Section~\ref{sec:fcn} derives functional sharp bounds for the Roy model with mixed discrete-continuous outcomes, and their implications for testability of Roy behavior and effects of endogenous sector selection on functional features such as inequality measures. 
Section~\ref{sec:appli} applies the derived bounds to the analysis of major choice in Canada and Germany and the under-representation of women in STEM. 
The last section concludes. Proofs of the main results are collected in the appendix.




\section{Analytical framework}
\label{sec:frame}

We adopt the framework of the potential
outcomes model $Y=Y_1D+Y_0(1-D),$ where $Y$ is an observed scalar outcome,
$D$ is an observed selection indicator, which takes value $1$ if Sector~1 is chosen, and $0$ if Sector $0$ is chosen, and $Y_1$, $Y_0$, are
unobserved potential outcomes, with common lower bound $\underline b$ on their support ($\underline b$ is usually~$0$ or~$-\infty$). \cite{HV:99} trace the
genealogy of this model and we refer to them for terminology and
attribution. The object of interest is the joint distribution of $(Y_0,Y_1)$ and features thereof.
Since $Y$ and $D$ are observed, the joint distribution of $(Y,D)$ is directly identified from the data.

We strip the model down to its self-selection mechanism, where agents are perfectly informed of the joint distribution of their potential outcomes $(Y_0,Y_1)$ in both sectors and choose the sector that maximizes outcomes, so that $D=1$ when $Y_1>Y_0$ and $D=0$ if $Y_1<Y_0$. The model is silent on the tie-breaking mechanism agents use in case $Y_1=Y_0$. As is customary in such frameworks, the assumption that agents are perfectly informed is intended to reflect, within a simple static model, the result of dynamic adjustments and learning on the one hand, and to put in stark relief the difference between the agents' and the analyst's information sets, on the other hand. 

We summarize the model with the following assumptions.
\begin{assumption}[Potential outcomes]
\label{ass:roy}
Observed outcomes are the realizations of a random variable $Y$ satisfying $Y=Y_1D+Y_0(1-D)$, where $(Y_0,Y_1)$ is a pair of possibly dependent unobserved random variables and $D$ is an observed indicator variable.
\end{assumption}
\begin{assumption}[Selection]
\label{ass:sel}
The selection indicator satisfies $Y_1>Y_0\Rightarrow D=1$, $Y_1<Y_0\Rightarrow D=0$.
\end{assumption}
Individuals choose the sector that yields higher outcome, when $Y_1\ne Y_0$. Their choice criterion is unspecified if $Y_1=Y_0$. When outcomes are discrete, the possibility of ties has to be considered. More generally, in a Roy model of earnings, the possibility of equal earnings in both sectors has to be entertained, if wage setters propose contracts that pool different skill levels, for instance. If the probability of ties is non zero, the Roy model specification described here is different from the specification of the competing risks model with non zero probability of ties in \cite{BM:97}. The model we consider here is tuned to economic applications, where the sector selection is unknown, when both sectors yield the same outcome. Hence, we identify $\mathbb P(Y_d\leq y,D=d)$, for $d=0,1,$ but not $\mathbb P(Y_d\leq y,Y_d>Y_{1-d})$. All we know is that $\mathbb P(Y_d\leq y,Y_d>Y_{1-d})\leq\mathbb P(Y_d\leq y,D=d)\leq\mathbb P(Y_d\leq y,Y_d\geq Y_{1-d})$. In the competing risks analysis of \cite{BM:97}, on the other hand, $\mathbb P(Y_1\leq y,Y_1>Y_0)$, $\mathbb P(Y_0\leq y,Y_1<Y_0)$ and $\mathbb P(Y_1\leq y,Y_1=Y_0)$ are all assumed identified, so that one observes when both components of the system fail simultaneously.

Our analysis can be extended to the case, where $Y$, $Y_0$ and $Y_1,$ take values in  an ordered subset of a Euclidean space, such as $\mathbb R^2$ endowed with the lexicographic order $\precsim_{{\text lex}}$, for instance. In the latter case, with outcome variable $Y=(W,T)$ ordered lexicographically, Assumption~\ref{ass:sel}, would read
$[ W_d>W_{1-d} \mbox{ or } [ W_d=W_{1-d} \mbox{ and } T_d > T_{1-d} ] ] \Rightarrow D=d.$ 
Take the case of university STEM major choice for instance. Lexicographic Roy preferences based on relatedness and income implies that prospective students choose STEM if they anticipate only STEM degrees will provide them with employment in their field of study, or if both STEM and non STEM provide them with employment in their field but they anticipate higher earnings in STEM. 
All results below would relate to the probability distributions of outcomes and potential outcomes relative to the chosen order, not the multivariate probability distributions. For instance, in the lexicographic example, the probability distribution is defined as $\mathbb P(Y\precsim_{{\text lex}}y)=\mathbb P((W,T)\precsim_{{\text lex}}(w,t))=\mathbb P(W<w\mbox{ or }[W=w\mbox{ and }T\leq t]).$

The whole analysis, model, distributional assumptions and theoretical results, are understood to be conditional on a set of observed covariates, which will be omitted from the notation, unless they are involved in identifying assumptions. 




\section{Binary outcome Roy model}
\label{sec:binary}

A great deal of the intuition for the characterization that we propose for the Roy model can be developed with the simplest version, where 
$Y_0$ and $Y_1$ are both binary outcomes. It models success or failure in securing a desired outcome, and the way it depends on a binary choice of treatment. 
In the case of college major choice, considered in Section~\ref{sec:appli}, $Y_1$ will model the ability to secure permanent employment at the time of the survey interview, if the degree or the major is classified as STEM, whereas $Y_0$ will model the ability to secure employment, with a non-STEM degree or major. 

\begin{definition}[Binary outcome Roy model]
\label{def:bin}
A model satisfying Assumptions~\ref{ass:roy} and~\ref{ass:sel}, with $Y_0,Y_1\in\{0,1\}$, is called {\em binary outcome Roy model}.
\end{definition}

An alternative way of defining a binary outcomes model, which shares the main features of the Roy model, i.e., self-selection on unobserved heterogeneity, involves {\em latent potential outcomes}. It is identical to the Roy model, except that potential outcomes are censored.
\begin{definition}[Alternative binary Roy model]
\label{def:alt}
Observed outcomes are the realizations of a random variable $Y$ satisfying $Y=Y_1D+Y_0(1-D)$, where
\begin{enumerate} 
\item potential outcomes satisfy $Y_d=1\{Y_d^\ast>0\}$, for $d=0,1,$ for a pair of possibly dependent unobserved random variables $(Y_0^\ast,Y_1^\ast)$,
\item $D$ is an observed indicator variable, satisfying $Y_1^\ast>Y_0^\ast\Rightarrow D=1$, $Y_1^\ast<Y_0^\ast\Rightarrow D=0$.
\end{enumerate}
\end{definition}
The alternative binary Roy model of Definition~\ref{def:alt}
can be interpreted in two ways. First, it is equivalent to a model with $Y=1\{Y^\ast>0\}$, where $Y^\ast$ satisfies a Roy model. Hence, it can be interpreted as a censored Roy model. The latent variables may be continuous variables, such as wages, and the analyst only observes whether or not they fall above or below a threshold. Other examples include examination grades, which are unobserved, except for the pass or fail outcome. Second, the actual outcome may be binary and be the result of a two-stage decision by the agent. In a first stage, they choose the sector of activity, with their choice of college major, for instance. In a second stage, they decide whether or not to work. The labor supply decision hinges on the difference between wage and reservation wage in the chosen sector. Then, $Y_d^\ast$ can be interpreted as the difference between wage in Sector $d$ and reservation wage in Sector $d$. If reservation wages are equal in both sectors, the model still conforms to the simple Roy incentive mechanism, where wages are the only determinant of sector choice. If reservation wages differ in both sectors, however, the model no longer conforms to the simple Roy incentive mechanism, as sector selection internalizes possibly non-pecuniary costs and benefits of each sector, as in the recent analysis of the generalized Roy model
in \cite{EHV:2014}.

Despite their distinct interpretations, it will be shown that sharp bounds for the joint distribution of $(Y_0,Y_1)$ are identical in both models, so that both models carry exactly the same information 
on the joint distribution of censored potential outcomes. They also share the reduced form implication
\begin{eqnarray}
\label{eq:att}
\mathbb E(Y_d-Y_{1-d}\vert D=d)\geq0,
\end{eqnarray}
which can be interpreted as a condition of {\em chosen sector advantage} or as nonnegative average treatment effect on the treated (where choice of Sector~1 corresponds to treatment). However, we show below that the reduced form condition~(\ref{eq:att}) contains less information on the joint distribution of potential outcomes than the structural models of Definitions~\ref{def:bin} and~\ref{def:alt} do. In particular, constraint~(\ref{eq:att}) is also shared by a binary outcome Roy model with imperfect foresight, identical to the binary outcome Roy model of Definition~\ref{def:bin}, except that the selection equation of Assumption~\ref{ass:sel} is replaced with the following:
\begin{assumption}(Imperfect foresight)
\label{ass:if}
The selection indicator satisfies 
$\mathbb E[Y_1-Y_0|\mathcal I]>0\Rightarrow D=1$, $\mathbb E[Y_0-Y_1|\mathcal I]>0\Rightarrow D=0$, where $\mathcal I$ is the sigma-algebra characterizing the agent's information set at the time of sector choice.
\end{assumption}
\begin{definition}[Binary outcome Roy with imperfect foresight]
\label{def:imp}
A model satisfying Assumptions~\ref{ass:roy} and~\ref{ass:if}, with $Y_0,Y_1\in\{0,1\}$, is called {\em binary outcome Roy model with imperfect foresight}.
\end{definition}
Our results in the next section characterize sharp bounds on the joint distributions of potential outcomes and highlight the difference in empirical content between perfect and imperfect foresight Roy models.


\subsection{Sharp bounds for the binary outcome Roy model}

In the binary outcomes Roy model, the lack of point identification comes from the fact that the mapping from observed sector and success to unobserved skills is not single valued. We know that when success in Sector~1 is observed, potential outcomes can be either $(Y_0=1,Y_1=1)$, i.e., success in both sectors, or $(Y_0=0,Y_1=1)$, i.e., success in Sector~1 only. Hence the identified probability that a random individual in the population chooses Sector~1 and succeeds will not be sufficient to identify the probability of succeeding in Sector~1. What we do know, however, is that 
$Y=0$ is observed if and only if the individual has neither the skills to succeed in Sector~0 nor in Sector~1. Hence,
$\mathbb P(Y_0=0,Y_1=0)= \mathbb P(Y=0)$.
Moreover, if the individual has the skills to succeed in Sector~0, but not in Sector~1, then, success in Sector~0 will be observed, so that
$\mathbb P(Y_0=1,Y_1=0)\leq\mathbb P(Y=1,D=0)$. Symmetrically, if the individual has the skills to succeed in Sector~1, but not in Sector~0, then, success in Sector~1 will be observed, so that
$\mathbb P(Y_0=0,Y_1=1)\leq\mathbb P(Y=1,D=1)$.

The discussion above shows that the expressions hold. Showing sharpness of these bounds is more involved, and the proof of the Proposition~\ref{prop:sharp} is given in the appendix,
together with a more fastidious statement of the theorem, with a rigorous and unambiguous definition of {\em sharp bounds} in this context. Note that the {\em bounds} can take the form of an equality in case upper and lower bounds coincide.
\begin{proposition}[Sharp bounds for the binary outcome Roy model]
\label{prop:sharp}
The following equality and inequalities provide a set of sharp bounds for the joint distribution of potential outcomes $(Y_0,Y_1)$ in the binary outcomes Roy model (Definition~\ref{def:bin}) and the alternative binary Roy model (Definition~\ref{def:alt}).
\begin{eqnarray}
\label{eq:sharp}
\begin{array}{lcl}
\mathbb P(Y_0=1,Y_1=0)&\leq&\mathbb P(Y=1,D=0),\\
\mathbb P(Y_0=0,Y_1=1)&\leq&\mathbb P(Y=1,D=1),\\
\mathbb P(Y_0=0,Y_1=0)&=& \mathbb P(Y=0).
\end{array}
\end{eqnarray}
\end{proposition}

The bounds in Proposition~\ref{prop:sharp} summarize all the information in the (alternative) binary outcome Roy model about the joint distribution of potential outcomes. From these bounds, sharp bounds on the marginals, which are akin to traditional bounds on average treatment outcomes, can be recovered. Combining the equality and inequalities of (\ref{eq:sharp}), we obtain traditional bounds on the marginals (see for example \cite{Manski:2007}, Section 7.5).
\begin{eqnarray}
\label{eq:marg}
\mathbb P(Y=1,D=0)\leq \mathbb EY_0 \leq \mathbb  P(Y=1)\mbox{ and } \mathbb P(Y=1,D=1)\leq \mathbb EY_1 \leq \mathbb  P(Y=1).
\end{eqnarray}
If the means of marginal potential outcomes are the objects of interest, as in \cite{Manski:90}, the bounds above are sharp without additional restrictions. Here, we take bounds on the joint distribution of potential outcomes as the object of interest. It is easy to see that (\ref{eq:marg}) and $\mathbb P(Y_0=0,Y_1=0)=\mathbb P(Y=0)$ are jointly equivalent to (\ref{eq:sharp}). However, from (\ref{eq:marg}) alone, (\ref{eq:sharp}) cannot be recovered, so that information on the joint distribution is lost. The bounds on the average sector difference are
\begin{eqnarray}
\label{eq:ate}
-\mathbb P(Y=1,D=0)\leq\mathbb E(Y_1-Y_0)\leq\mathbb P(Y=1,D=1).
\end{eqnarray}
The sharp bounds of Proposition~\ref{prop:sharp} emphasize two important facts:
\begin{enumerate}
\item On the one hand, despite the literature on {\em non identification} of competing risks, starting with \cite{Cox:72} and \cite{Tsiatis:75}, the Roy model does in fact contain non trivial information about the joint distribution of potential outcomes, hence of skills, or more generally, of sector specific unobserved heterogeneity. 
\item On the other hand, the sharp bounds of Proposition~\ref{prop:sharp} can be very wide and they do not cross. For any joint distribution for $(Y,D)$, there exists a joint distribution for $(Y_0,Y_1)$ that fits the binary outcome Roy model, so that the latter is not falsifiable in the absence of additional constraints.
\end{enumerate}
Since the Roy model imposes strong restrictions on behavior, the lack of testability is particularly vexing. We shall consider exclusion and monotonicity restrictions that allow us to recover testability of behavior characterized by Roy sector selection. In the case of college major choice, considered in Section~\ref{sec:appli}, one of our main concerns will be with explanations of the under representation of women in STEM. One candidate is wage discrimination in STEM, which is compatible with a Roy model of behavior. Another is gender profiling in major choice, which is not. Hence the ability to test Roy maximizing behavior in major selection is paramount.


\subsection{Stochastically monotone instrumental variables (SMIV)}

In order to allow falsifiability of the Roy model, we now investigate the implications of exclusion restrictions. Such exclusions are of two types: sector-specific variables, i.e., variables affecting only one outcome equation, but not the other (Assumption~\ref{ass:ex} in Appendix~\ref{app:cov}), and variables that shift sector selection, but shift potential outcomes either not at all (Assumption~\ref{ass:IV} below), or only in one direction (Assumption~\ref{ass:SMIV} below). To sharpen the focus and save space, we discuss the conceptually relatively straightforward implications of sector specific variables in Appendix~\ref{app:proofs}, and consider mostly the effect of vectors $Z$ of variables that affect sector selection, but have restricted impact on potential outcomes. We shall comment on the way in which sector specific exclusions modify the expressions and leave details to Appendix~\ref{app:cov} (As mentioned before, conditioning on remaining observed covariates is implicit in all the paper).
We start the discussion with variables that shift selection, but not potential outcomes.
\begin{assumption}
\label{ass:IV}
There exists a vector $Z$ of observable random variables, such that $(Y_0,Y_1)\perp\!\!\!\perp Z$.
\end{assumption}
Such variables are akin to typical instrumental variables, and examples within Roy models in the existing literature include parental education in \cite{WR:79}, distance to a college in \cite{EHV:2014} and attendance in a Catholic high school in \cite{AET:2005}. Local aggregate labor market variables at the time of sector selection are also often used, as in \cite{EHV:2014} and references therein.

First, it is important to emphasize, that, unlike the generalized Roy model used in the contributions cited in the previous paragraph, the pure Roy selection mechanism imposes $D=1$ when $Y_1>Y_0$ and $D=0$ when $Y_1<Y_0$. Hence, a selection shifter $Z$ satisfying Assumption~\ref{ass:IV} can only affect the model in case of ties $Y_1=Y_0$. The model is lexicographic, in the sense that agents care only about outcomes when choosing their sector of activity, unless the outcomes are equal in the two sectors, at which point other considerations guide their decision. As a result, $Y$ is independent of $Z$, but $(Y,D)$ is not jointly independent of $Z$, so that the bounds in Proposition~\ref{prop:sharp} can be sharpened using variation in $\mathbb P(Y=1,D=1\vert Z)$ and in $\mathbb P(Y=1,D=0\vert Z)$. Taking the expressions in (\ref{eq:sharp}) conditionally on $Z$ and using Assumption~\ref{ass:IV} to remove conditioning in the left-hand sides yields the bounds
{ \scriptsize
\begin{eqnarray}
\label{eq:sharp-z}
\begin{array}{lcl}
\mathbb P(Y_0=1,Y_1=0)&\leq&\inf_z\mathbb P(Y=1,D=0\vert Z=z),\\\\
\mathbb P(Y_0=0,Y_1=1)&\leq&\inf_z\mathbb P(Y=1,D=1\vert Z=z),\\\\
\mathbb P(Y_0=0,Y_1=0)&=& \mathbb P(Y=0)\;=\;\mathbb P(Y=0\vert Z=z).
\end{array}
\end{eqnarray}
}
The tightened bounds are proven to be sharp in Appendix~\ref{app:proofs} and illustrated in Figure~\ref{fig:2simplex}.
They are {\em intersection bounds}, and inference can be carried out with the method proposed in \cite{CLR:2009}.

The third expression in (\ref{eq:sharp-z}) gives a testable implication, since the binary outcomes Roy model under Assumptions~\ref{ass:IV} implies $Y\perp\!\!\!\perp Z$. We now argue that the latter summarizes all possible testable implications of the model. Indeed, for any joint distribution of $(Y,D,Z)$ on $\{0,1\}^2\times\mathbb R$ satisfying $Y\perp\!\!\!\perp Z$, we can always define the pair of potential outcomes $(Y_0,Y_1)$ by $Y_0=Y_1:=Y$ and satisfy the constraints of the binary outcome Roy model
under Assumption~\ref{ass:IV}.

However, rejection of $Y\perp\!\!\!\perp Z$ cannot be attributed to a violation of the Roy selection assumption (Assumption~\ref{ass:sel}) if the validity of the instrument is under question. 
In the case of college major choice, considered in Section~\ref{sec:appli}, one of the proposed instrument is parental education. Unfortunately, the validity of this instrument is doubtful, as parental education level may be correlated with unobserved individual productivity in one or both of the sectors, hence affect potential outcomes directly.
Indeed, \cite{CHS:2010} argue that cognitive and non-cognitive unobserved skills are determined in great part by parental environment and investment, which in turn is highly correlated with parental education. Distance to college is a similarly tainted instrument for returns to education, as discussed in \cite{Card:2001}, since parental location preferences are correlated with unobserved cognitive and non cognitive parental investments. The same applies to local labor market conditions, which may drive endogenous location choices.
Moreover, \cite{KM:2015} derive sharp testable implications of $(Y_0,Y_1)\perp\!\!\!\perp Z$ within a binary potential outcomes model (without the Roy selection assumption) and their test tends to reject validity of parental education as an instrument, including in our data. More generally, instruments are elusive in the study of returns to education.
The rest of this section is concerned with a weakening of Assumption~\ref{ass:IV} and a discussion of its validity, in order to recover testable implications of the Roy selection assumption.

Our objective now is to bring covariate information to bear and restore falsifiability of the Roy selection mechanism without relying on strong independence assumptions that are hard to substantiate. Joint independence of potential labor market outcomes and parental education is indeed hard to substantiate, as unobserved benefits of parental education can raise productivity. However, it is natural to assume that increasing parental education cannot worsen potential labor market outcomes (see the discussion after the statement of Assumption~\ref{ass:SMIV}). Similarly, local aggregate labor market variables, such as the average wage in STEM for an individual socio-economic category at the time of college major decision, are also likely to be correlated with ex-post job market outcomes, but higher local average wages in STEM at the time of major decision are unlikely to produce lower wages in STEM at the time of graduation, barring complex general equilibrium adjustments. Measures of school quality, merit based scholarships, and distance to college also fall in the category of useful variation shifters that are typically not independent of potential outcomes, but may shift them in only one direction.

The following weakening of Assumption~\ref{ass:IV} formalizes this insight. We adopt the following notion of monotonicity for the instrument. For details, refer to \cite{SS:2007}, Section~6B. When comparing vectors, ``$\geq$'' denotes the componentwise partial order.
\begin{definition}
\label{def:FSD}(First Order Stochastic Dominance)
A distribution $F_1$ on $\mathbb R^k$ is said to be {\em first order stochastically dominated} by a distribution $F_2$ if there exists random vectors $Y_1$ with distribution $F_1$ and $Y_2$ with distribution $F_2$ such that $\mathbb P(Y_2\geq Y_1)=1$. By extension, a random vector with distribution $F_2$ is also said to stochastically dominate a random vector with distribution $F_1$.
\end{definition}

\begin{assumption}(SMIV)
\label{ass:SMIV}
For any pair $z_2\geq z_1$ in the support of a vector of observable variables~$Z$, the conditional distribution of $(Y_0,Y_1)$ given $Z=z_2$ first order stochastically dominates the distribution of $(Y_0,Y_1)$ given $Z=z_1$ (denoted $(Y_0,Y_1)\vert Z=z_2\succsim_{FSD}(Y_0,Y_1)\vert Z=z_1$).
\end{assumption}

Assumption~\ref{ass:SMIV} is inspired by the monotone instrumental variable (hereafter MIV) of \cite{MP:2000}. Assumption~\ref{ass:SMIV} is stronger than MIV, which, adapted to our setting, would only constrain the means of potential outcomes, but not the marginal distributions or any feature of the joint distribution, which are crucial to the sharp bounds and the testing procedure we develop.\footnote{\scriptsize Note, however, that observed variable~$\nu$ (researcher measured ability), presented as an example of MIV in Lemma~3.1 of \cite{MP:2009}, can be shown to actually satisfy our SMIV Assumption~\ref{ass:SMIV} under the assumptions of the lemma.} Assumption~\ref{ass:SMIV} is different from Constraint~(9) in \cite{BGIM:2007}, which restricts the endogenous realized wage distribution, whereas our assumption is meant to operate on the vector of skills. As we shall discuss below, in our framework, Constraint (9) in BGIM is a testable implication of wage maximization behavior under Assumption~\ref{ass:SMIV}.
Note also that Assumption~\ref{ass:SMIV} can hold with respect to a vector of instruments, which can increase the tightness of bounds on parameters of interest.

We now discuss the economic content and validity of Assumption~\ref{ass:SMIV} within the context of a sector selection model. Assume, as in the skill formation technology of \cite{CH:2007} \cite{CH:2008} and \cite{CHS:2010}, that the vector of potential outcomes, such as the potential wages in both sectors, is determined by $(Y_0,Y_1)=f(\theta,p,\eta)=(f_0(\theta,p_0,\eta_0),f_1(\theta,p_1,\eta_1))$, where~$\theta=(\theta_C,\theta_N)$ is a vector of cognitive and non cognitive skills (or abilities), $p_d$, $d=0,1,$ is a vector of prices of cognitive and non cognitive skills in Sector~$d$, $\eta_d$, $d=0,1,$ is a shock in Sector~$d$, and~$f_d$, $d=0,1,$ is a scalar function (see Equations~(7) and~(9) of \cite{HHV:2018} for a special case). Suppose the function~$f$ is increasing in~$\theta$, which is a reasonable assumption, even if different sectors value cognitive and non cognitive skills differently. If~$(\theta_C,\theta_N)$ is stochastically monotone with respect to a (vector of) determinant(s)~$Z$ of skill investment, and~$Z$ is independent of prices and shocks, then, the vector of potential outcomes~$(Y_0,Y_1)$ will inherit from~$(\theta_C,\theta_N)$ stochastic monotonicity with respect to~$Z$, and Assumption~\ref{ass:SMIV} will hold.


\subsection{Sharp bounds and testability of the binary outcomes Roy model under SMIV}\label{sec:test}

An important distinction between the roles of the independence assumption (Assumption~\ref{ass:IV}) and the stochastic monotonicity assumption (Assumption~\ref{ass:SMIV}) is that, under the former, the instrument $Z$ can only shift sector selection when $Y_0=Y_1$, whereas under the latter, $Z$ is no longer required to be independent of potential outcomes~$(Y_0,Y_1)$ and can therefore induce variation in $D$, even when~$Y_0\ne Y_1$.

To see how the stochastic monotonicity assumption (Assumption~\ref{ass:SMIV}) combines with the Roy selection mechanism (Assumption~\ref{ass:sel}), start from the sharp bounds of Proposition~\ref{prop:sharp} in the equivalent representation 
{ \scriptsize
\begin{eqnarray*}
\begin{array}{ccccc}
\mathbb P(Y=1,D=0\vert Z=z)&\leq&\mathbb P(Y_0=1\vert Z=z)&\leq&\mathbb P(Y=1\vert Z=z),\\
\mathbb P(Y=1,D=1\vert Z=z)&\leq&\mathbb P(Y_1=1\vert Z=z)&\leq&\mathbb P(Y=1\vert Z=z),\\
\mathbb P(Y_0=Y_1=0\vert Z=z)&=&\mathbb P(Y=0\vert Z=z).
\end{array}
\end{eqnarray*}
}
The statement
$(Y_0,Y_1)\vert Z=z_2\succsim_{FSD}(Y_0,Y_1)\vert Z=z_1$
is equivalent to
$\mathbb P((Y_0,Y_1)\in U\vert Z=z_2)\geq\mathbb P((Y_0,Y_1)\in U\vert Z=z_1),$
for all upper sets $U$ (Theorem 6.B.1 of \cite{SS:2007}, Section~6B).
\begin{definition}(Upper Sets)
\label{def:US}
A subset $U$ of a partially ordered set $(\mathcal S,\geq)$ is called an {\em upper set} if $y\in U$ implies $\tilde y\in U$ for all $\tilde y\geq y$.
\end{definition} 
The non trivial upper subsets of  $\{0,1\}^2$ are \[\{(1,1)\}, \{(0,1),(1,1)\}, \{(1,0),(1,1)\}, \{(0,1),(1,0),(1,1)\}.\] Consider, for instance, the upper set $\{(0,1),(1,1)\}$. Stochastic monotonicity of $(Y_0,Y_1)$ in $z$ implies that
$\mathbb P((Y_0,Y_1)\in \{(0,1),(1,1)\}\vert Z=z)\leq\mathbb P((Y_0,Y_1)\in \{(0,1),(1,1)\}\vert Z=\tilde z)$ for all $\tilde z \geq z$, or equivalently $\mathbb P(Y_1=1\vert Z=z)\leq\mathbb P(Y_1=1\vert Z=\tilde z)$. Since the latter is smaller than or equal to $\mathbb P(Y=1\vert Z=\tilde z)$ by Assumptions~\ref{ass:roy} and~\ref{ass:sel}, we obtain
$\mathbb P(Y_0=1\vert Z=z)\leq\mathbb P(Y=1\vert Z=\tilde z)$ for all~$\tilde z\geq z$ in the domain of~$Z$. Proceeding similarly with all upper subsets of~$\{0,1\}$, we obtain
the following sharp bounds for the joint distribution of potential outcomes under Assumptions~\ref{ass:roy},~\ref{ass:sel} and~\ref{ass:SMIV}: for all~$z$ in the domain of~$Z$,
{ \scriptsize
\begin{eqnarray}
\label{eq:SMIV}
\begin{array}{ccccc}
\sup_{\tilde z\leq z}\mathbb P(Y=1,D=0\vert Z=\tilde z)&\leq&\mathbb P(Y_0=1\vert Z=z)&\leq&\inf_{\tilde z\geq z}\mathbb P(Y=1\vert Z=\tilde z),\\
\sup_{\tilde z\leq z}\mathbb P(Y=1,D=1\vert Z=\tilde z)&\leq&\mathbb P(Y_1=1\vert Z=z)&\leq&\inf_{\tilde z\geq z}\mathbb P(Y=1\vert Z=\tilde z),\\
\sup_{\tilde z\geq z}\mathbb P(Y=0\vert Z=\tilde z)&\leq&\mathbb P(Y_0=Y_1=0\vert Z=z)&\leq&\inf_{\tilde z\leq z}\mathbb P(Y=0\vert Z=\tilde z).
\end{array}
\end{eqnarray}
}
The third line of the display in (\ref{eq:SMIV}) combines identification of $\mathbb P(Y_0=Y_1=0\vert Z=z)$, which is equal to $\mathbb P(Y=0\vert Z=z)$, and the testable implications $\sup_{\tilde z\geq z}\mathbb P(Y=0\vert Z=\tilde z)\leq\inf_{\tilde z\leq z}\mathbb P(Y=0\vert Z=\tilde z)$ for all $z$ in the domain of~$Z$. The latter is equivalent to stochastic monotonicity of $Y$ in $z$, which turns out to summarize all testable implications of Roy under Assumption~\ref{ass:SMIV} as formalized in the following theorem.

\begin{theorem}[Sharp bounds and testable implications of Roy under SMIV]
\label{thm:SMIV} 

\hskip2pt{  }\hskip2pt

\begin{enumerate}
\item The display in (\ref{eq:SMIV}) characterizes the identified set for the joint distribution of potential outcomes in the binary outcomes Roy model under Assumptions~\ref{ass:roy}, \ref{ass:sel} and~\ref{ass:SMIV}.
\item Under Assumptions~\ref{ass:roy},~\ref{ass:sel} and~\ref{ass:SMIV}, the following holds: ($\ast$) For any pair $z_2\geq z_1$ in the support of the vector of observable variables~$Z$, $Y\vert Z=z_2\succsim_{FSD}Y\vert Z=z_1$.
\item If $(Y,Z)$ satisfies ($\ast$), then there is a pair $(Y_0,Y_1)$ such that Assumptions~\ref{ass:roy},~\ref{ass:sel} and~\ref{ass:SMIV} hold.
\end{enumerate}
\end{theorem}

When $Y$ is stochastically monotone in $z$, $\inf_{\tilde z\geq z}\mathbb P(Y=1\vert Z=\tilde z)$ is equal to $\mathbb P(Y=1\vert Z=z)$, which, by the third line of (\ref{eq:SMIV}) is equal to $1-\mathbb P(Y_0=Y_1=0\vert Z=z)$. Hence the right-hand side inequalities in the first two lines of (\ref{eq:SMIV}) are redundant, and the identified set for the joint distribution of potential outcomes is characterized by two inequalities and one equality. The proof of Theorem~\ref{thm:SMIV}(1) is given in Appendix~\ref{app:proofs} and the identified set is represented graphically  on the right-hand-side panel of Figure~\ref{fig:2simplex}. 
The proof of Theorem~\ref{thm:SMIV}(2,3) is straightforward. Indeed, under Assumptions~\ref{ass:roy} and~\ref{ass:sel}, we have $\mathbb P(Y_0\leq y,Y_1\leq y\vert Z)=\mathbb P(Y\leq y\vert Z)$, for all $y$, since counterfactual outcomes cannot be larger than realized ones. Hence stochastic monotonicity of $(Y_0,Y_1)$ immediately implies stochastic monotonicity of $Y$. We now argue that it constitutes a sharp testable implication of the Roy selection mechanism. Indeed, given any joint distribution of observable variables $(Y,D,Z)$ on $\mathbb R\times\{0,1\}\times\mathbb R^d$, with $Y\vert Z=z_2\succsim_{FSD}Y\vert Z=z_1$ for each $z_2\geq z_1$, the pair of potential outcomes $(Y_0,Y_1)$ can always be chosen in such a way that Assumptions~\ref{ass:roy}, \ref{ass:sel} and~\ref{ass:SMIV} are satisfied. For example, setting $Y_0=Y_1=Y$ would satisfy all the constraints.

Theorem~\ref{thm:SMIV} shows that testing the Roy selection mechanism simply boils down to testing stochastic monotonicity of observed outcomes with respect to the monotone instrumental variable, which can be performed with existing inference methods in \cite{LLW:2009}, \cite{DE:2012} and \cite{HLS:2016}. Statements~(2) and~(3) of Theorem~\ref{thm:SMIV} make no mention of the binary outcomes Roy model, since they are valid without restrictions on the domain of the outcome variables. Theorem~\ref{thm:SMIV} also sheds new light on Assumption~(9) in \cite{BGIM:2007}, which is identical to our testable implication of the Roy model, when the outcome of interest is wage. Hence, the stochastic monotonicity constraint of~\cite{BGIM:2007} can be seen as an implication of wage maximization behavior in the sector selection stage.


\subsection{Imperfect foresight}

To address the concern that rejection of the Roy selection mechanism may be down to rejecting the assumption that agents are perfectly informed of their future potential outcomes at the time of sector selection, we also derive testable implications of the binary outcome Roy model with imperfect foresight of Definition~\ref{def:imp}. The latter is identical to the binary outcome Roy model, except that the Roy selection assumption,  Assumption~\ref{ass:sel}, is replaced by imperfect foresight, namely Assumption~\ref{ass:if}. 

Under the potential outcomes model, i.e., Assumption~\ref{ass:roy}, only, we still know that an individual with the skill to succeed in Sector~0, but not in Sector~1, will be observed as having succeeded in Sector~0 or as having failed in Sector~1 (the latter was ruled out under the Roy selection rule of Assumption~\ref{ass:sel}). Hence $\mathbb P(Y_0=1,Y_1=0\vert Z)\leq\mathbb P(Y=1,D=0\vert Z)+\mathbb P(Y=0,D=1\vert Z)$ and symmetrically for $\mathbb P(Y_0=0,Y_1=1\vert Z)$. An individual without the skills to succeed in either sector will be observed to fail, so that $\mathbb P(Y_0=0,Y_1=0\vert Z=z)\leq\mathbb P(Y=0\vert Z=z)$. Under stochastic monotonicity in $Z$ (Assumption~\ref{ass:SMIV}), the latter yields $\mathbb P(Y_0=0,Y_1=0\vert Z=z)\leq\inf_{\tilde z\leq z}\mathbb P(Y=0\vert Z=\tilde z)$.  In addition, observing success in Sector~$d$ necessary implies that the agent has the skills required for Sector~$d$, hence $\mathbb P(Y_d=1\vert Z)\geq\mathbb P(Y=1,D=d\vert Z)$. Under Assumption~\ref{ass:SMIV}, the latter yields $\mathbb P(Y_d=1\vert Z=z)\geq\sup_{\tilde z\leq z}\mathbb P(Y=1,D=d\vert Z=\tilde z),$ $d\in\{0,1\}$. 

We now add selection information according to Assumption~\ref{ass:if}. The latter is equivalent to \[Y=Y_d\Rightarrow \mathbb E[Y\vert\mathcal I]=\mathbb E[Y_d\vert\mathcal I]\geq\mathbb E[Y_{1-d}\vert\mathcal I], \; d=0,1.\] This yields $\mathbb E(Y\vert \mathcal I)=\max\{\mathbb E(Y_0\vert \mathcal I),\mathbb E(Y_1\vert \mathcal I)\}$. Under Assumption~\ref{ass:SMIV}, the latter yields monotonicity of $\mathbb E(Y\vert Z=z)$ in $z$. Note that the same testable implications would be obtained had Assumption~\ref{ass:if} been replaced with a more general sector selection rule based on the comparison of expected utilities, namely the rule $\mathbb E[u(Y_d)\vert\mathcal I]>\mathbb E[u(Y_{1-d})\vert \mathcal I]\Rightarrow D=d$, for $d=0,1.$

Putting it all together yields the following sharp bounds on the joint distribution of potential outcomes under Assumptions~\ref{ass:roy},~\ref{ass:if} and~\ref{ass:SMIV}:
{ \scriptsize
\begin{eqnarray}
\label{eq:gen-if}
\begin{array}{lllll}
\mathbb P(Y_0=1,Y_1=0\vert Z=z)&\leq&\mathbb P(Y=1,D=0\vert Z=z)&+&\mathbb P(Y=0,D=1\vert Z=z),\\
\mathbb P(Y_0=0,Y_1=1\vert Z=z)&\leq&\mathbb P(Y=0,D=0\vert Z=z)&+&\mathbb P(Y=1,D=1\vert Z=z),\\
\mathbb P(Y_0=0,Y_1=0\vert Z=z)&\leq&1-\mathbb E(Y\vert Z=z),
\end{array}
\end{eqnarray}
and
\begin{eqnarray}
\label{eq:marg-if}
\begin{array}{cll}
\sup_{\tilde z\leq z}\mathbb P(Y=1,D=0\vert Z=\tilde z)&\leq&\mathbb E(Y_0\vert Z=z),\\
\sup_{\tilde z\leq z}\mathbb P(Y=1,D=1\vert Z=\tilde z)&\leq&\mathbb E(Y_1\vert Z=z),\\
\max\{\mathbb E(Y_0\vert Z=z),\mathbb E(Y_1\vert Z=z)\}&=&\mathbb E[Y\vert Z=z],
\end{array}
\end{eqnarray} 
}
for all $z$ in the support of $Z$.

The inequalities above define the identified set for the joint distribution of potential outcomes. Testable implications of the Roy model with imperfect foresight include monotonicity of $\mathbb E[Y\vert Z=z]$ in $z$ as derived above, which proves Theorem~\ref{thm:if}(2) below. It can be easily shown that in the binary case, this monotonicity summarizes the empirical content of the Roy selection assumption with imperfect foresight, as stated in Theorem~\ref{thm:if}. Indeed, for any given vector $(Y,D,Z)$ such that $\mathbb E[Y\vert Z=z]$ is non decreasing in~$z$, setting $Y_0=Y_1=Y$ satisfies the assumptions, which proves Theorem~\ref{thm:if}(3) below.

\begin{theorem}[Testable implications of Roy with imperfect foresight]
\label{thm:if}

\hskip2pt{  }\hskip2pt

\begin{enumerate}
\item The displays in~(\ref{eq:gen-if}) and~(\ref{eq:marg-if}) jointly characterize the identified set for the joint distribution of potential outcomes in the binary outcomes Roy model with imperfect foresight under Assumptions~\ref{ass:roy}, \ref{ass:if} and~\ref{ass:SMIV}.
\item If Assumptions~\ref{ass:roy},~\ref{ass:if},~\ref{ass:SMIV} hold with $\mathcal{I}$-measurable~$Z$, then $\mathbb E[Y\vert Z=z]$ is non decreasing in~$z$.
\item For any distribution $G$ on $\{0,1\}^2\times\text{Supp}(Z)$, such that $\mathbb E[Y\vert Z=z]$ is non decreasing in~$z$, there exists a random vector $(Y_0,Y_1,D,Z)\in\{0,1\}^3\times\text{Supp}(Z)$ such that $(Y_1D+Y_0(1-D),D,Z)$ has distribution~$G$ and Assumptions~\ref{ass:if} and~\ref{ass:SMIV} are satisfied with $\mathcal I=\sigma(Z)$. 
\end{enumerate}
\end{theorem}
We can therefore test Roy with imperfect foresight under Assumption~\ref{ass:SMIV} simply by testing monotonicity of $\mathbb E[Y\vert Z=z]$ in~$z$, using existing inference methods in \cite{Chetverikov:2013} or \cite{HLS:2016}. We can also verify that stochastic monotonicity of~$Y$ (the testable implication of Roy selection as shown in Theorem~\ref{thm:SMIV}) does indeed imply monotonicity of $\mathbb E[Y\vert Z=z]$ in~$z$, which is consistent with the fact that Assumption~\ref{ass:sel} implies Assumption~\ref{ass:if}. Moreover, in the binary outcomes case, the testable implications of Roy behavior under SMIV and those of imperfect foresight Roy under SMIV are identical, since when $Y$ is binary, stochastic monotonicity of $Y\vert Z=z$ and monotonicity of $\mathbb E[Y\vert Z=z]$ are equivalent. Hence, rejection of Roy selection behavior under SMIV implies rejection of Roy with imperfect foresight as well. However, the identified set for the joint distribution of potential outcomes in Theorem~\ref{thm:SMIV}(1) is nested in, and weakly tighter than the identified set of Theorem~\ref{thm:if}(1), since the combination of Assumptions~\ref{ass:sel} and~\ref{ass:SMIV} contains more information on the joint distribution of potential outcomes than the combination of Assumptions~\ref{ass:if} and~\ref{ass:SMIV}.


\subsection{Bounds on departures from Roy selection}\label{sec:dep}

In case of rejection of the Roy selection mechanism, the methodology developed here, and particularly the information on the joint distribution of potential outcomes, allows us to quantify departures from the Roy sector selection rule (Assumption~\ref{ass:sel}). If agents are believed to be expected outcome maximizers, i.e., to behave according to the binary outcome Roy model with imperfect foresight, this measure of departure can be interpreted as a measure of the cost of imperfect foresight. If, on the other hand, departures from the Roy model with imperfect foresight are entertained, then the measure of departure we propose also captures the extent to which considerations other than potential outcome maximization enter in the decision. This may be the result of maximization of a utility function that depends on aspects beyond the chosen outcomes. It may also reveal a bias in decision making. This would be the case, in our application to major choice, if gender profiling discouraged women from choosing STEM majors. 

Departure from Roy sector selection, which we therefore interpret as inefficiency of sector choice, can be measured as the difference between maximum potential outcome and realized outcome, the two being equal by definition in the case of Roy selection according to Assumption~\ref{ass:sel}. 
\begin{definition}[Efficiency loss]
\label{def:el}
Efficiency loss from Roy selection departures is defined for each $z\in$ Supp$(Z)$ as el$(z):=\mathbb P(\max(Y_0,Y_1)=1\vert Z=z)-\mathbb P(Y=1\vert Z=z)$ in the binary outcomes case, and, for each $y\in$ Supp$(Y)$, as el$(y,z):=\mathbb P(Y\leq y\vert Z=z)-\mathbb P(\max(Y_0,Y_1)\leq y\vert Z=z)$, otherwise. 
\end{definition}
We have $\mathbb P(\max(Y_0,Y_1)=1\vert Z=z)-\mathbb P(Y=1\vert Z=z)=\mathbb P(Y=0\vert Z=z)-\mathbb P(Y_0=Y_1=0\vert Z=z)$. Since in the binary outcomes Roy model, $\mathbb P(Y_0=Y_1=0\vert Z=z)$ is identified as $\mathbb P(Y=0\vert Z=z)$, efficiency loss is zero, which justifies the interpretation as a departure from Roy selection (Assumption~\ref{ass:sel}). When Assumption~\ref{ass:sel} is dropped, efficiency loss is non negative. Since $\mathbb P(Y=0\vert Z=z)$ is identified, bounds on efficiency loss or departure from Roy will be obtained from bounds on $\mathbb P(Y_0=Y_1=0\vert Z=z)$ under Assumptions~\ref{ass:roy} and~\ref{ass:SMIV} only. Since $\mathbb P(Y_0=Y_1=0\vert Z=z)$ involves the joint distribution of potential outcomes, sharp bounds on marginal distributions alone cannot deliver the desired bounds on efficiency loss. This feature is shared by other policy relevant parameters such as ex-post regret, skill correlation, sector effect conditional on the chosen sector and the proportion who benefit from a given sector, all of which can also be bounded using this methodology.

Note that the efficiency loss criterion is instrument-dependent. First, it is a function of the value~$z$ of the instrument~$Z$, which is part of the appeal, since it allows to identify regions of the support of the instrument that are most susceptible to departures from wage maximization. Second, after taking the expectation over~$Z$, the parameter no longer depends on the choice of instrument, however, the bounds do. 
If there are two instruments and the vector also satisfies SMIV, then the whole analysis can be carried out with respect to the vector of instruments to tighten the bounds.

Under Assumptions~\ref{ass:roy} and~\ref{ass:SMIV}, the identified set for the joint distribution of potential outcomes is obtained in a similar fashion to~(\ref{eq:gen-if})-(\ref{eq:marg-if}), except that we cannot rely on selection information, so that the upper bounds in~(\ref{eq:marg-if}) are obtained from the fact that talent for Sector~$d$ only precludes observing failure in Sector~$d$. Bounds~(\ref{eq:gen-if})-(\ref{eq:marg-if}) are therefore replaced by
{ \scriptsize
\begin{eqnarray}
\label{eq:gen-smiv}
\begin{array}{lllll}
\mathbb P(Y_0=0,Y_1=0\vert Z=z)&\leq&\inf_{\tilde z\leq z}\mathbb P(Y=0\vert Z=\tilde z)&&\\
\mathbb P(Y_0=0,Y_1=1\vert Z=z)&\leq&\mathbb P(Y=0,D=0\vert Z=z)&+&\mathbb P(Y=1,D=1\vert Z=z),\\
\mathbb P(Y_0=1,Y_1=0\vert Z=z)&\leq&\mathbb P(Y=1,D=0\vert Z=z)&+&\mathbb P(Y=0,D=1\vert Z=z),\\
\mathbb P(Y_0=1,Y_1=1\vert Z=z)&\leq&\inf_{\tilde z\geq z}\mathbb P(Y=1\vert Z=\tilde z),&&
\end{array}
\end{eqnarray}
and
\begin{eqnarray}
\label{eq:marg-smiv}
\begin{array}{lllll}
\sup_{\tilde z\leq z}\mathbb P(Y=1,D=0\vert Z=\tilde z)&\leq&\mathbb P(Y_0=1\vert Z=z)&\leq&1-\sup_{\tilde z\geq z}\mathbb P(Y=0,D=0\vert Z=\tilde z)\\\\
\sup_{\tilde z\leq z}\mathbb P(Y=1,D=1\vert Z=\tilde z)&\leq&\mathbb P(Y_1=1\vert Z=z)&\leq&1-\sup_{\tilde z\geq z}\mathbb P(Y=0,D=1\vert Z=\tilde z),
\end{array}
\end{eqnarray} for all $z$ in the support of $Z$. 
}
From~(\ref{eq:gen-smiv}), we obtain immediately an upper bound on~$\mathbb P(Y_0=Y_1=0\vert z)$, namely $\inf_{\tilde z\leq z}\mathbb P(Y=0\vert Z=\tilde z)$. Sharp bounds are obtained by projecting~(\ref{eq:gen-smiv})-(\ref{eq:marg-smiv}) onto component $\mathbb P(Y_0=Y_1=0\vert z)$ in the 3-simplex.

\begin{proposition}[Bounds on efficiency loss]
\label{prop:el}
\hskip1pt{ }\hskip1pt
\begin{enumerate}
\item The displays in~(\ref{eq:gen-smiv}) and~(\ref{eq:marg-smiv}) jointly characterize the identified set for the joint distribution of potential outcomes under Assumptions~\ref{ass:roy} and~\ref{ass:SMIV}, with $Y\in\{0,1\}$.
\item Under Assumptions~\ref{ass:roy} and~\ref{ass:SMIV} with $Y\in\{0,1\}$, efficiency loss due to departures from Roy selection satisfies, for each {\scriptsize $z\in$ Supp$(Z)$,
$el(z)\;\geq\;\mathbb P(Y=0\vert Z=z)-\inf_{\tilde z\leq z}\mathbb P(Y=0\vert Z=\tilde z).$}
\end{enumerate}
\end{proposition}
The results on efficiency loss allow us to identify values of observable characteristics that induce the most costly misallocation of talent and field of study and are therefore prime targets for intervention.




\section{Roy model with discrete-continuous outcomes}
\label{sec:fcn}

Extending the analysis to richer sets of outcomes, including mixed discrete and continuous potential outcomes does not remove the lack of identification issue in the Roy model (and the related competing risks model). The range of observables is richer, but so is the object of interest, i.e., the joint distribution of potential outcomes. Given partial observability and endogenous sector selection, the Roy model is essentially partially identified. Results obtained in the form of sharp bounds on the joint distribution of potential outcomes and the methods used to derive them are analogous to the corresponding results and methods in the binary outcome case, except in one important respect. When considering distributional aspects, such as inequality, the distinction between {\em pointwise bounds} and {\em functional bounds} is crucial as described below. After a discussion of the latter point, we proceed to analyze testability and quantifying departures from the Roy selection mechanism along the same lines as in the binary outcomes case in Section~\ref{sec:binary}.

\subsection{Functionally sharp bounds for the Roy model}

Consider the Roy model of Section~\ref{sec:frame}, under Assumptions~\ref{ass:roy} and~\ref{ass:sel}. Bounds on the marginal distributions of potential outcomes can be derived very easily as follows. 
For any real number $y$,
$\mathbb P(Y_d\leq y)=\mathbb P(Y_d\leq y, D=d)+\mathbb P(Y_d\leq y, D=1-d).$
The first term on the right-hand-side is identified. The second term on the right-hand-side is bounded below by
$\mathbb P(Y_{1-d}\leq y, D=1-d)$, and above by $\mathbb P(D=1-d).$ The resulting bounds
were shown by \cite{Peterson:76} to be {\em pointwise sharp} for the marginal distributions of potential outcomes, in the sense that any pair of distributions of potential outcomes that satisfy the bounds for a given fixed $y$, can be obtained from some joint distribution of observable variables $(Y,D)$ under the assumptions of the Roy model.
However, as \cite{Crowder:91} pointed out, there are additional non redundant cross quantile restrictions, namely, for all $y\geq x$,
$\mathbb P(x<Y_d\leq y)\geq\mathbb P(x< Y_d\leq y, D=d)$.
If the object of interest involves densities, such as the hazard rate, or functional features, such as inequality measures, the difference between the latter bounds and pointwise bounds can be considerable. Indeed, combining Peterson bounds 
involves an additional term $-\mathbb P(D=1-d)$ in the lower bound. This difference arises because the monotonicity of the distribution function is not factored in.
Graphically, the difference between pointwise bounds and functional bounds can be highlighted on Figure~\ref{fig:iqr}. A candidate distribution function for $Y_d$ that is drawn through the two points $(\tilde y_1,q_1)$ and $(\tilde y_2,q_2)$ can lie between the curves $\mathbb P(Y\leq y)$ and $\mathbb P(Y\leq y,D=d)+\mathbb P(D=1-d)$. Hence it satisfies pointwise bounds. However, its slope is lower in some regions than the slope of the curve $\mathbb P(Y\leq y,D=d)$, so that it fails to satisfy the functional bounds.

Turning to the joint distribution function of potential outcomes, pointwise bounds can also be derived very easily. Indeed, we immediately have
\begin{eqnarray}
\label{eq:pjoint}
\begin{array}{l}
\mathbb P(Y\leq \min(y_0,y_1))\hskip5pt\leq\hskip5pt
\mathbb P(Y_0\leq y_0,Y_1\leq y_1)
\hskip5pt\leq\hskip5pt\mathbb P(Y\leq y_0,D=0)+\mathbb P(Y\leq y_1,D=1).
\end{array}
\end{eqnarray}
Corollary~1 of \cite{BM:97} 
shows that the bounds (\ref{eq:pjoint}) can be attained under their competing risks specification. However, once again, these bounds fail to incorporate monotonicity conditions, and they can entail loss of information, when describing functional features of potential outcomes.

The object of interest is the joint distribution $(Y_0,Y_1)$, the information on which we wish to characterize using the identified joint distribution of observable variables $(Y,D)$.
Take any subset $A$ of $\mathbb R^2$ and consider bounding the probability of $(Y_0,Y_1)\in A$. If $A$ contains points $(y_0,y)$ such that $y_0\leq y$, it can give rise to observation $(Y=y,D=1)$, and if $A$ contains points $(y,y_1)$ such that $y_1\leq y$, it can give rise to observation $(Y=y,D=0)$. Hence, observation $(Y=y,D=d)$ such that $y\in U_{A,0}$ below, and only those, can be rationalized by elements of $A$. Similarly, to derive the lower bound, notice that $(Y,D)=(y,1)$ can arise for any $(Y_0,Y_1)\in[\underline b,y]\times\{y\}$, so that $(Y_0,Y_1)$ mass could be concentrated outside $A$ unless the whole of $[\underline b,y]\times\{y\}$ is contained in $A$ (where $\underline b$ is the common lower bound of the supports of $Y_0$ and~$Y_1$).
\begin{definition}\label{def:up-low}
For any Borel set $A$ in $\mathbb R^2$, define the sets $U_{A,0}$, $U_{A,1}$ and $L_{A,0}$, $L_{A,1}$ as
\begin{eqnarray*}
\begin{array}{ll}
U_{A,0}=\{y\in\mathbb R\;|\;\{y\}\times[\underline b,y]\cap A\ne\varnothing\},&
L_{A,0}=\{y\in\mathbb R\;|\;\{y\}\times[\underline b,y]\subseteq A\},\\
U_{A,1}=\{y\in\mathbb R\;|\;[\underline b,y]\times\{y\}\cap A\ne\varnothing\},&
L_{A,1}=\{y\in\mathbb R\;|\;[\underline b,y]\times\{y\}\subseteq A\}.
\end{array}
\end{eqnarray*}
\end{definition}
We shall formally show that the upper bound is $\mathbb P(Y\in U_{A,0},D=0)+\mathbb P(Y\in U_{A,1},D=1)$. Similarly, the lower bound will be shown to be $\mathbb P(Y\in L_{A,0},D=0)+\mathbb P(Y\in L_{A,1},D=1)$. In the case, where $A$ is an upper set (Definition~\ref{def:US}), the bounding sets of Definition~\ref{def:up-low} take a very simple form, and we derive the constraints associated with Assumption~\ref{ass:SMIV} (SMIV) accordingly.

\begin{theorem}[Sharp bounds for the joint distribution]
\label{thm:func-sharp}
\hskip1pt{ }\hskip1pt
\begin{enumerate}
\item Let the distribution of observable variables $(Y,D)$ on $\mathbb R\times\{0,1\}$ be given. Under Assumptions~\ref{ass:roy} and~\ref{ass:sel}, the distribution of $(Y_0,Y_1)$ on $\mathbb R^2$ satisfies, for all Borel subset $A$ of $\mathbb R^2$,
{ \scriptsize
\begin{eqnarray*}
&&\mathbb P(Y\in L_{A,0},D=0)+\mathbb P(Y\in L_{A,1},D=1)\\
&&\hskip50pt\leq\;\mathbb P((Y_0,Y_1)\in A)\\
&&\hskip100pt\leq\mathbb P(Y\in U_{A,0},D=0)+\mathbb P(Y\in U_{A,1},D=1).
\end{eqnarray*}
}
\item Conversely, for any joint distribution satisfying the bounds above, there exists a pair $(Y_0,Y_1)$ with that distribution, which satisfies Assumptions~\ref{ass:roy} and~\ref{ass:sel}.
\item If Assumption~\ref{ass:SMIV} also holds, then the distribution of $(Y_0,Y_1)$ also satisfies, for all upper set $A$ of $\mathbb R^2$, all $z\in$ Supp$(Z)$,
{ \scriptsize
\begin{eqnarray*}
&&\sup_{\tilde z\leq z}[\mathbb P(Y\geq\underline y_0^A,D=0\vert Z=\tilde z)+\mathbb P(Y\geq\underline y_1^A,D=1\vert Z=\tilde z)]\\
&&\hskip150pt\leq\;\mathbb P((Y_0,Y_1)\in A\vert Z=z)\;\leq\;\inf_{\tilde z\geq z}\mathbb P(Y\geq\underline y^A\vert Z=\tilde z),
\end{eqnarray*}
with $\underline y^A:=\inf\{y:(y,y)\in A\},$ $\underline y_0^A:=\inf\{y: (y,+\infty)\times\mathbb R\subseteq A\}$ and $\underline y_1^A:=\inf\{y: \mathbb R\times(y,+\infty)\subseteq A\}$.
}
\end{enumerate}
\end{theorem}

Theorem~\ref{thm:func-sharp}(1) allows us to easily recover Peterson
bounds with suitable choices of $A$. Choosing $A=[\underline b,y]\times\mathbb R$ yields Peterson bounds on the marginal distribution of $Y_0$. Choosing $A=[\underline b,y_0]\times[\underline b,y_1]$ yields Peterson bounds on the joint distribution of $(Y_0,Y_1)$. 
Finally, applying Theorem~\ref{thm:func-sharp} to sets of the form $(y_{1},y_{2}]\times\mathbb R$ and $\mathbb R\times(y_{1},y_{2}]$ yields the following bounds on the marginal distributions of $Y_d$, for $d=1,0$:
{ \scriptsize
\begin{eqnarray}
&&\mathbb P(y_{1}<Y\leq y_{2},D=d)+\mathbb P(Y\leq y_{2}, D=1-d)1\{y_1\leq\underline b\}
\label{eq:low}\\
&&\hskip50pt\leq \mathbb P(y_{1}<Y_d\leq y_{2})
\nonumber\\
&&\hskip100pt\leq\mathbb P(y_{1}<Y\leq y_{2},D=d)+\mathbb P(y_{1}<Y, D=1-d).
\label{eq:up}
\end{eqnarray}
}
The upper bound (\ref{eq:up}) is redundant. Indeed, it can be recovered from lower bounds on $\mathbb P(y_{2}<Y_d\leq\infty)$ and $\mathbb P(\underline b<Y_d\leq y_{1})$. We shall show that the class of sets of the form $(y_{1},y_{2}]\times\mathbb R$ and $\mathbb R\times(y_{1},y_{2}]$ suffice to characterize the marginal potential distributions and that the lower bounds are functionally sharp, as formulated in Corollary~\ref{cor:marg} below. The bounds are similar, though not identical, to the bounds in Theorem~1 of \cite{BM:97} for a related competing risks model, discussed in the paragraph below Assumption~\ref{ass:sel}. The result is proved in the appendix, with a more rigorous statement and formal definition of {\em functional sharp bounds}.
\begin{corollary}[Sharp bounds for the marginal distributions]\label{cor:marg}
Under Assumptions~\ref{ass:roy} and~\ref{ass:sel}, the bounds 
\[\mathbb P(y_{1}<Y_d\leq y_{2})\geq
\mathbb P(y_{1}<Y\leq y_{2},D=d)+\mathbb P(Y\leq y_2, D=1-d)1\{y_1\leq\underline b\}
\] 
for all $y_1, y_2\in\mathbb R\cup\{\pm\infty\}$, $y_1<y_2$, and $d=0,1$, 
are functional sharp bounds.
\end{corollary}
Corollary~\ref{cor:marg} tells us that intervals are sufficient to characterize all the information we have on the marginal distribution of potential outcomes $(Y_1,Y_0)$. They form a {\em core determining} class of sets, in the terminology of \cite{GH:2006}, \cite{GH:2011}.
This has several advantages. It allows the incorporation of exclusion restrictions and lends itself to the partial identified inference of \cite{CLR:2009} and \cite{AS:2011}. The characterization of Corollary~\ref{cor:marg} allows us to derive sharp bounds on functional features such as measures of inequality.

\subsection{Testing Roy and bounding departures from Roy}\label{sec:depF}
As in the binary outcome case of Section~\ref{sec:binary}, the Roy model defined by Assumptions~\ref{ass:roy} and~\ref{ass:sel}, is not falsifiable without additional information. Indeed, for any joint distribution $(Y,D)$, potential outcomes $(Y_0,Y_1)$ can be chosen, for instance with $Y_0=Y_1=Y$, such that Assumptions~\ref{ass:roy} and~\ref{ass:sel} hold. Given the unavailability of an instrument that satisfies Assumption~\ref{ass:IV}, we examine falsifiability of the model under the stochastically monotone instrumental variable assumption (Assumption~\ref{ass:SMIV}). Theorem~\ref{thm:SMIV}(2,3) shows that stochastic monotonicity of observed outcomes with respect to the instrument summarizes all observable implications of the Roy selection mechanism under SMIV. Hence Roy selection behavior can be tested using existing inferential methods to test stochastic monotonicity. As concerns falsifiability of the Roy model with imperfect foresight, Theorem~\ref{thm:if}(2) shows that a testable implication is monotonicity of~$\mathbb E[Y\vert Z=z]$ in~$z$, which can also be tested using existing inference methods on regression monotonicity. However, Theorem~\ref{thm:if}(3) only holds in the binary outcomes case, since monotonicity of $\mathbb E[Y\vert Z=z]$ does not otherwise imply stochastic monotonicity of~$Y\vert Z=z$ in~$z$, and therefore does not summarize the empirical content of Roy with imperfect foresight under Assumption~\ref{ass:SMIV}.

According to Definition~\ref{def:el}, departure from Roy selection behavior or inefficiency of sector choice can be measured with the difference $\mathbb P(Y\leq y\vert Z=z)-\mathbb P(\max(Y_0,Y_1)\leq y\vert Z=z)$. The latter is zero under Assumption~\ref{ass:sel} (Roy selection mechanism). Otherwise, $\mathbb P(Y\leq y\vert Z=z)-\mathbb P(\max(Y_0,Y_1)\leq y\vert Z=z)\geq\mathbb P(Y\leq y\vert Z=z)-\inf_{\tilde z\leq z}\mathbb P(Y\leq y\vert Z=\tilde z)$, under Assumption~\ref{ass:SMIV}. 
\begin{proposition}
\label{prop:el-gen}
Under Assumptions~\ref{ass:roy} and~\ref{ass:SMIV}, efficiency loss of Definition~\ref{def:el} satisfies 
\[
{\text el}(y,z)\;\geq\;\mathbb P(Y\leq y\vert Z=z)-\inf_{\tilde z\leq z}\mathbb P(Y\leq y\vert Z=\tilde z),
\]
for all $(y,z)$ in the support of $(Y,Z)$.
\end{proposition}
As the binary case, the lower bound on efficiency loss is zero under Assumption~\ref{ass:sel} and can serve to construct a test statistic for a test of Roy selection behavior.

\subsection{Functional features of potential distributions}
The original motivation of the Roy model was to analyze the effect of self-selection on wage distributions, and particularly on wage inequality.
\cite{HH:90} show that self-selection reduces aggregate inequality when skills are log normal and within sector inequality when skills have a log concave distribution. One of the purposes of functional sharp bounds derived in the previous section is to analyze the effect of self-selection on inequality of potential outcomes in the specification of the Roy model we consider here, where the Roy model structure is stripped down to the self-selection mechanism. Functional sharp bounds on the potential outcome distributions allow us to derive sharp bounds on inequality measures. In this section, we concentrate on the interquantile range, although the same reasoning applies to other functionals from the vast literature on distributional inequality.

Consider two quantiles $q_1$ and $q_2$ with $q_2>q_1$, as illustrated on Figure~\ref{fig:iqr}. The most commonly used range is the interquartile range, where $q_1=1-q_2=1/4$, but other cases, such as $q_1=1-q_2=0.1,$ are also of great empirical relevance. Peterson bounds on the distribution of $Y_d$ impose $\mathbb P(Y_d\leq y_1)\leq \mathbb P(Y\leq y_1,D=d)+\mathbb P(D=1-d)=q_1$ and $\mathbb P(Y_d\leq y_2)\geq\mathbb P(Y\leq y_2)=q_2$. Hence, the upper bound on the interquantile range based on pointwise sharp bounds for the distribution of $Y_d$ is $y_2-y_1$. However, functional sharp bounds of Corollary~\ref{cor:marg} are violated, since $q_2-q_1<\mathbb P(y_1<Y\leq y_2,D=d)$. On Figure~4, we exhibit another pair of points, namely $(\tilde y_1,q_1)$ and $(\tilde y_2,q_2)$ such that a distribution for potential $Y_d$ cannot cross these two points and satisfy the functional sharp bounds of Corollary~\ref{cor:marg}.
\begin{figure}[htbp]
\begin{center}
\caption{{\scriptsize Sharp bounds on the interquantile range for the distribution of $Y_d$. The pointwise upper bound for the range between quantiles $q_1$ and $q_2$ is $y_2-y_1$. However, range $\tilde y_2-\tilde y_1$ violates functional sharp bounds because $q_2-q_1<\tilde q_2-\tilde q_1$.}}
\label{fig:iqr}
\vskip15pt
\includegraphics[width=0.9\textwidth]{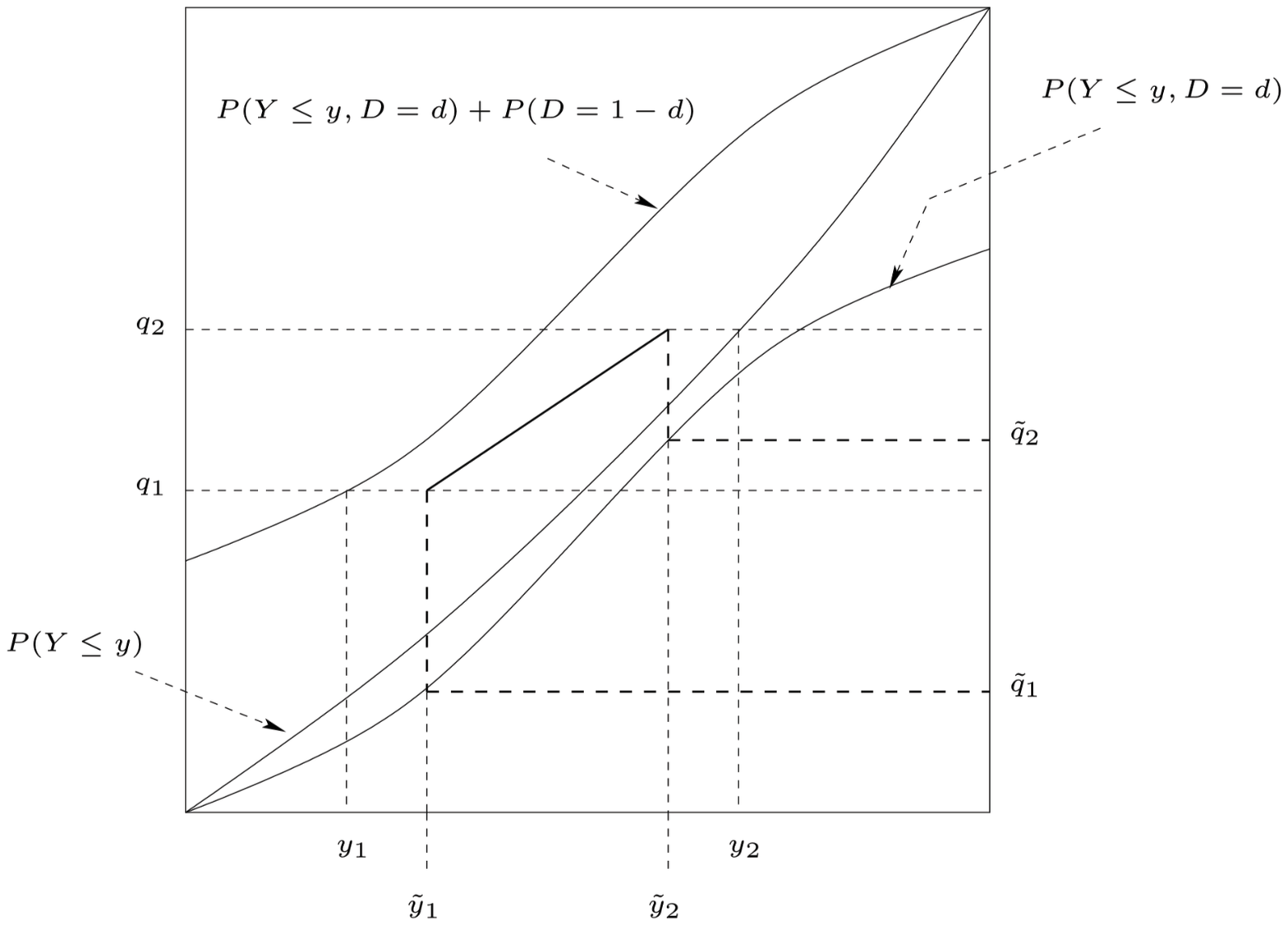}
\end{center}
\end{figure}

We now show how to derive sharp bounds for the interquantile range. For ease of notation throughout this section, for $d=0,1$, and for each $y\in\mathbb R$, denote
$F(y):=\mathbb P(Y\leq y),$
$F_d(y):=\mathbb P(Y_d\leq y),$
$\underbar F_d(y):=\mathbb P(Y\leq y, D=d),$
$\bar F_d(y):=\mathbb P(Y\leq y,D=d)+\mathbb P(D=1-d),$
and $f^{-1}$ the generalized inverse of $f$, i.e., $f^{-1}(q)=\inf\{y:f(y)>q\}$.
Start from any $y$ within the pointwise quantile bounds
$\bar F_d^{-1}(q_1)\leq y\leq F^{-1}(q_2).$
From $y$, the largest interquantile range obtains in either of the following two cases:
\begin{enumerate}
\item when $F(y)$ is hit first, in which case the interquantile range is $F^{-1}(q_2)-y$,
\item when the potential distribution $F_d$ follows the slope of $\underbar F_d$ starting from the point with coordinates $(y,q_1)$, in which case the interquantile range is 
$\tilde y-y$, where $\tilde y$ achieves 
\[
\sup\{\tilde y:\;q_2\geq  q_1+\underbar F_d(\tilde y)-\underbar F_d(y)\}.
\]  
\end{enumerate}
Hence, the interquantile range starting from quantile $y$ is:
\[
\mbox{IQR}(y)=\min\left( F^{-1}(q_2)-y, \underbar F_d^{-1}(q_2-q_1+\underbar F_d(y))-y\right).
\] 
Finally, maximizing IQR($y$) over admissible $y$'s yields the upper bound on the interquantile range.
Hence, under Assumptions~\ref{ass:roy} and \ref{ass:sel}, the sharp bounds on the interquantile $(q_1,q_2)$ range are given~by:
\begin{eqnarray}
\begin{array}{l}
\max(0,\bar F_d^{-1}(q_2)-F^{-1}(q_1)) \hskip10pt\leq\hskip10pt\mbox{IQR}(q_1,q_2) \hskip10pt\leq\hskip10pt\\\\
\hskip75pt 
\max_{\bar F_d^{-1}(q_1)\leq y\leq F^{-1}(q_1)}
\left(
\min\left( F^{-1}(q_2)-y, \underbar F_d^{-1}(q_2-q_1+\underbar F_d(y))-y\right)
\right). 
\end{array}
\label{eq:iqr}
\end{eqnarray}
Under Assumption~\ref{ass:SMIV} (SMIV), bounds on the interquantile range would be constructed in the same way, conditionally on $Z$, with {\scriptsize $F(y)$} replaced with {\scriptsize $\sup_{\tilde z\leq z}\mathbb P(Y\leq y\vert Z=\tilde z)$} and {\scriptsize $\bar F_d(y)$} replaced with {\scriptsize $\inf_{\tilde z\leq z}[\mathbb P(Y\leq y,D=d\vert Z=\tilde z)+\mathbb P(D=1-d\vert Z=\tilde z)]$}.

From the viewpoint of the interquantile range, we can now consider the effect of self-selection into the sector of activity (or treatment) on inequality, both within sector and in the aggregate. We compare outcome distributions resulting from self-selection, hereafter called outcome distributions in the {\em self-selection economy}, to distributions of outcomes that would result from random assignment of individuals to sectors of activity, hereafter called outcome distributions in the {\em random assignment economy}, as in \cite{HS:85}, \cite{HS:90} and \cite{HH:90}. In Sector $d$, the distribution of outcomes in the random assignment economy is the distribution of potential outcome $Y_d$, while the distribution of outcomes of the self-selection economy is $\mathbb P(Y\leq y\vert D=d)$. In the aggregate population, the distribution of outcomes of the random assignment economy is $\mathbb P(Y_0\leq y)\mathbb P(D=0)+\mathbb P(Y_1\leq y)\mathbb P(D=1)$, whereas the distribution of outcomes of the self-selection economy is simply the distribution of observable outcomes $Y$. These cases are collected in Table~\ref{table:iqr}.

\begin{table}
{\scriptsize
\caption{Distribution of outcomes under self-selection and random assignment.}
\label{table:iqr}
\begin{center}
\begin{tabular}{l|cc}
&&\\
&self-selection economy&random assignment economy\\&&\\\hline\\
Sector $d$&$\mathbb P(Y\leq y\vert D=d)$&$\mathbb P(Y_d\leq y)$\\\\
Aggregate&$\mathbb P(Y\leq y)$
&$\mathbb P(Y_0\leq y)\mathbb P(D=0)+\mathbb P(Y_1\leq y)\mathbb P(D=1)$\\&&
\end{tabular}
\end{center}
}
\end{table}

In Sector $d$, the interquantile range between quantiles $q_1$ and $q_2$ of the distribution of outcomes in the random assignment economy is bounded above by (\ref{eq:iqr}). In the self-selection economy, it is identified as the interquantile range of the distribution $\mathbb P(Y\leq y\vert D=d)$. The following proposition shows how they compare.
\begin{proposition}[Inequality in Sector $d$]
\label{prop:ineq}
$ $
\begin{enumerate}
\item If the distribution of outcomes $Y$ conditional on $D=d$ first order stochastically dominates the distribution of outcomes $Y$ conditional on $D=1-d$, i.e., $\mathbb P(Y\leq y\vert D=d)\leq\mathbb P(Y\leq y\vert D=1-d)$ for all $y\in\mathbb R$, then, for any pair of quantiles, the interquantile range of the distribution of outcomes in Sector~$d$ in the self-selected economy is lower than the upper bound of the interquantile range of the distribution of outcomes in Sector~$d$ in the random assignment economy.
\item If the stochastic dominance relation of (1) does not hold, then there exists distributions for the pair $(Y,D)$ such that the interquantile range of the distribution of outcomes in Sector~$d$ in the self-selected economy is larger than the upper bound of the interquantile range of the distribution of outcomes in Sector~$d$ in the random assignment economy.
\end{enumerate}
\end{proposition}
Proposition~\ref{prop:ineq} tells us two things. On the one hand, if Sector~$d$ is unambiguously more profitable in the self-selected economy, it is possible for inequality in Sector $d$, as measured by the interquantile range, to decrease with self-selection, relative to an economy with random assignment of individuals to sectors. On the other hand, if neither sector dominates the other in the self-selection economy, then there are joint distributions of observables under which we know that self-selection unambiguously increases inequality in Sector $d$. In case no sector stochastically dominates the other, the hypothesis that self-selection increases inequality is testable based on the bounds of (\ref{eq:iqr}), in the sense that one can test the hypothesis that the interquantile range in the self-selected economy is larger than the upper bound of the interquantile range in the randomized economy.




\section{Roy model of college major choice in Canada and Germany}
\label{sec:appli}

Since \cite{JACT:89} pointed out that major choice mattered more to labor market outcomes than college choice, the literature on returns to college education has placed some focus on the determinants of major choice and the effects on labor market outcomes. The salient classification that has come to dominate the debate is between STEM and non-STEM degrees, and there is ample evidence of the labor market advantages conferred on male graduates by STEM degrees: \cite{JACT:89} and \cite{Arcidiacono:2004} for the US, \cite{kelly2010} for Ireland, \cite{chevalier2011} for the UK, \cite{maselli2014} for several EU countries. 

The wage benefits of STEM degrees have been found to be a significant but not sole determinant of major choice. \cite{Arcidiacono:2004} finds that high ability students view education as a consumption good in the US. \cite{beffy2012} find elasticity of major choice to expected income to be significant, but less important as a determinant of major choice in France than heterogeneity in preferences for the subject matter. We revisit the issue using our nonparametric bounds methodology on Canadian and German data. We examine whether the data is consistent with a Roy selection of students into the two sectors based on anticipated labor market outcomes only. We study how the answers depend on visible minority status and residency in Qu\'ebec and the former German Democratic Republic. 

The picture is rather different for women. The labor market advantages, if present, are not so clear-cut, as noted by \cite{zafar2013} and \cite{HGHM:2013}, and women are severely under-represented in STEM education and even more so in STEM jobs. The evidence is summarized in \cite{beede2011}.
Two dominant explanations for the under-representation of women in STEM education and in STEM careers are discrimination, which lowers expected wages for women in STEM, and gender profiling, which keeps young women away from STEM education. The former is compatible with a Roy model of career choice, assuming wage discrimination is anticipated, and can be addressed by policies fighting lower labor market outcomes for women in STEM.
The latter involves non pecuniary considerations in major choice, therefore requires generalized Roy modeling and can be addressed by policies aimed at encouraging young women into STEM education. However, differential costs between STEM and non STEM majors are nonexistent in Germany and Canada, so that generalized Roy models based on differential costs are not directly applicable here.
In any event, given the divergence in policy implications of the two channels above, it is important to investigate which of the two is the dominant effect.

The under-representation of women in STEM jobs is often cited as a major contributor to the gender wage gap, as in \cite{daymont1984}. More generally, there is a large amount of informal discussion, although, to the best of our knowledge, little formal investigation, of the contribution of the STEM economy to rising wage inequality; see, for instance, \cite{BM:2012}, who attribute rising inequality to skill-based technological change. Our methodology allows us to address this issue by comparing inequality in STEM wages to inequality in non-STEM wages in a counterfactual economy, where sector allocation is random. We can also investigate the effect of self-selection on sectoral and aggregate wage inequality.

\subsection*{Data}

Our empirical analysis relies on Canadian and German nationally representative graduates surveys. Both countries have a tradition of running extensive surveys on graduate education, and they differ substantially on the proportions of graduates choosing STEM fields ($53\%$ of our German sample versus~$20\%$ of our Canadian sample), so that analyzing both simultaneously allows us to present a more robust picture of the differences in choice of men and women in graduate education. Both data sets contain detailed information on a representative sample of recent university graduates in their respective countries. The German data are collected by the German Centre for Higher Education Research and Science Studies (DZHW) as part of the DZHW Graduate Survey Series. Data and methodology are described in \cite{DZHW}. In Germany, the wave we consider includes graduates who obtained their highest degree during the academic year 2008-2009. The Canadian data is drawn from the National Graduate Survey of Statistics Canada. In Canada, the wave we consider includes graduates who obtained their highest degree during the academic year 2009-2010. We also examine data from earlier waves, namely 1997, 2001, 2005 for Germany and 2000, 2005 for Canada. In the case of the earlier Canadian waves, we rely on publicly available data, which has fewer variables than the data we use for the 2009-2010 wave, and which, unlike the latter, only provides interval censored income information. 

Graduates were interviewed 1 year and 5 years after graduation in the German survey and 3 years after graduation in the Canadian survey. At that point, extensive information was collected on their educational experience, employment history, including wages and hours worked, along with detailed socio-economic variables. Geographical information is more precise in the German data, with 38 regions, as opposed to 13 in the Canadian data. The German data also contains information on talent, with results at the {\em Abitur} (high school final exam), whereas the Canadian survey only provides a self-assessed measure of ability.

Both data sets allow us to observe whether employment is permanent or temporary and whether it is related to the specific field of study.
In both data sets, fields of study are recorded at a high level of disaggregation, which allows us to discriminate subjects that require mathematics from those that don't. We then merge the fields of study into two categories. We call STEM the mathematics intensive category, which consists mostly of mathematic, physical, economic and computer sciences, as well as engineering and related fields, although other STEM definitions often include life sciences and exclude economics. The remaining majors are merged in the non-STEM-degree category. In Canada and in Germany, the choice of field of study is made prior to enrolment in the program. In both countries, we only consider graduates from institutions in the country of the survey, who are active on their respective country's labor market at the time of the interview.

We consider a selection of outcome variables: the ability to secure a permanent employment, the ability to secure employment within the field of study, and annual wage and average hourly wage during the year prior to the time of the interview. Given the high correlation between wage and hourly wage measures, we report only results for wage. Annual wage is non censored in the Canadian data and reported in $1,000$ euro bins in the German data.

The potential instruments we consider are the education level of both parents (the surveys report parental education in discrete categories, which we translate into years of education following \cite{Parey:2017}), the proportion of women among STEM faculty members (which we call rate of feminization of the STEM faculty)
in universities in the individual's region of residence at the time of choice. 
The German version of this variable is drawn from data on gender distribution of faculty by field and by federal State provided by the Federal Statistical Office of Germany (DESTATIS).
The Canadian version of this variable is drawn from Statistics Canada, University and College Academic Staff System (UCASS). There is a very high level of assortative matching in parent's education both in Canada and Germany, so we only report results using the mother's education and the rate of feminization of STEM faculty as stochastically monotone instrumental variables. We also use local labor market conditions at the time of choice as instruments for robustness purposes, although their validity relies on neglecting general equilibrium effects.

We compare results for gender and visible minority status. In the Canadian survey, visible minority status is self-reported. In Germany, we construct this variable from the country of birth, and we assign an individual in the survey the status of visible minority if they were born in a country with a non-white majority population. This unfortunately excludes a large number of graduates of Turkish descent, whom we are unable to track. We also distinguish German graduates from institutions in the former German Democratic Republic and Canadian graduates from institutions in Qu\'ebec.

Our study focuses on the latest cohort. The raw sample from the German survey consists of 10,494 individuals.
From the raw sample, we exclude all respondents who are still in education, have never worked or are currently inactive, unemployed, in part-time employment or self-employed. This leaves 9,202 observations. We keep only graduates who hold a ``Bachelor'', ``Magister'' or ``Diplom'', excluding those with ``Staatsexamen'' and ``Lehramt'' degrees, which are specific tracks mainly for teachers. This leaves us with 7,729 observations. Finally, we divide the population between those who completed the \textit{Abitur} (high school final exam) in the former Federal Republic and in the former Democratic Republic and exclude those for which we do not have this information or obtained their Abitur abroad (107 individuals).
Most of our econometric analysis is based on the sample of individual with complete information on gender, degree, migration background, year and place of Abitur completion, mother's education, and income or job characteristics, that is between 4,559 and 4,890 observations.

The raw sample from the Canadian survey consists of 28,715 observations who participated in the survey. From the raw sample, we exclude all respondents who have completed trade, vocational, college and CEGEP diploma or certificate at the time of their 2009/2010 graduation and 2013 interview. We keep only those individuals who have ``university diploma or certificate below Bachelor level'', ``Bachelor’s degree or first professional degree'', ``university diploma/certificate above the Bachelor’s level but below the Master’s level'', ``Master’s degree'' and ``Doctorate''. We also exclude all respondents who are still in education, self-employed, working in family business without pay and live in the U.S. as primary residence. We further filter the data set to include respondents who are in the labor force, employed, work full-time and have age below 40. Since the econometric analysis is based on the sample with complete information on gender, minority status, income, degree, related job, permanent job, mother’s education and father’s education, the sample size ranges between 4,361 and 10,150 observations.

\subsection*{Descriptive statistics}

Income distributions in Germany and Canada show a clear STEM advantage for both men and women and a clear gender gap. In Figure~\ref{fig:Quartiles}, distributions appear to be stochastically ordered. In both Canada and Germany, based on quartiles only, the distribution of male STEM wages dominates the distribution of female STEM wages, which dominates male non STEM wages, which dominates female non STEM wages. 
A similar pattern emerges from Table~\ref{table:Perm}, where we see that men with STEM degrees are more likely to hold permanent employment in a field related to their studies, than men with non STEM degrees and women in both categories. More precisely, in Germany, $41\%$ of men with STEM degrees obtain permanent employment one year after graduation, $35\%$  in a field related to their studies and $6\%$ in other fields. For women with STEM degrees, the proportion is only $36\%$, with $29\%$ in their field of study, and for men and women with non STEM degrees, the proportion falls to $23\%$, with $16\%$ in their field of study. 
In Canada, $90\%$ of men with STEM degrees obtain permanent employment three years after graduation, $84\%$  in a field related to their studies and $6\%$ in other fields. For women with STEM degrees, the proportion is $82\%$, with $73\%$ in their field of study, and for men and women with non STEM degrees, the proportion is $81\%$, with $68\%$ in their field of study.
Since the proportion of men with STEM degrees is larger, the overall proportion of women with a permanent employment after 1 year in Germany is lower ($27\%$) than for men ($36\%$) and the proportion of women with a permanent employment after 3 years in Canada is lower (80\%) than for men (86\%). 

Table~\ref{table:STEM} shows the degree of under representation of women in STEM degrees in both Germany and Canada, which tallies with the overwhelming evidence from previous studies in different contexts. In Germany, $37\%$ of women's degrees are in STEM, as opposed to $75\%$ for men. The difference is somewhat less pronounced for minorities, where $48\%$ of women's degrees are in STEM, as opposed to $80\%$ for men. In Canada, $8\%$ of women's degrees are in STEM, as opposed to $35\%$ for men. The difference is, again, less pronounced for minorities, where $15\%$ of women's degrees are in STEM, as opposed to $45\%$ for men.

We examine the variation in sector choice induced by the instruments and illustrate it in the case of white women in affluent regions in Figure~\ref{fig:INS}, where the brown line is the point estimator and the grey lines are the~$95\%$ confidence bands. There is some indication of a hump-shaped response of STEM choices in mother's education.  The humped-shaped response to mother's education may be due to a larger involvement in major choices for parents with a bachelor's degree and a more laissez-faire approach beyond that. The effect of the proportion of women on the STEM faculty on women's choices is increasing for low proportions, as we would expect, then levels for larger proportions. 

\subsection*{Discussion of the SMIV instruments}

We contend that female mentors and maternal educational attainment are two determinants of skill investment that affect the vector of cognitive and non cognitive skills monotonically. Take maternal educational attainment first. A large literature on the intergenerational transmission of human capital suggests that cognitive skills are positively impacted by parental education (see for instance \cite{BDS:2005}). More recent evidence points to the same conclusion about non cognitive skills (see for instance \cite{Currie:2009} and \cite{CMP:2013}). Combined with evidence of complementarity between cognitive and non cognitive skills (\cite{CHS:2010}, \cite{HM:2014} and references therein), this tends to support stochastic monotonicity  of the vector of cognitive and non cognitive skills with respect to maternal educational attainment, hence the validity of Assumption~\ref{ass:SMIV} for the latter variable. Consider now the presence of female mentors on the faculty, or more precisely the proportion of women on the STEM faculty. \cite{BL:2005} report that ``theory and evidence suggest that female instructors may be instrumental in encouraging women to enroll and excel in subjects in which they are underrepresented.'' Hence, we expect that female instructors will help female students improve and adapt their cognitive and non cognitive skills to the demands of the market, hence increasing the vector of potential outcomes, so that Assumption~\ref{ass:SMIV} is satisfied.

\subsection*{Methodology and results}

From the survey samples, we first construct sub-samples based on gender, visible minority status, and the broad region of residence at the time of the interview (former East and West Germany, Qu\'ebec and the rest of Canada). We are interested in comparing behavior by gender and by race or immigrant status, as well as socio-economic background. As far as the latter determinant is concerned, since we have no data on socio-economic background of respondents, we use a coarse subdivision in regions for each of the countries, differentiating the relatively poorer former East Germany and Qu\'ebec. The latter division has the added benefit of distinguishing very different cultural spheres, the role of which we can also investigate.

We test monotonicity of the conditional mean for each binary outcome, and both mean monotonicity and stochastic monotonicity of the non-binary discrete and continuous outcomes with respect to the instruments. We implement the stochastic monotonicity test proposed in \cite{HLS:2016}.\footnote{\scriptsize We thank Yu-Chin Hsu and  Chu-An Liu for sharing their code.} The sensitivity of inference results to the generalized moment selection procedure is usually the major concern with this type of procedure, see for instance \cite{CS:2018}\footnote{\scriptsize The generalized moment selection procedure, originally introduced in \cite{Hansen:2005}, \cite{GH:2009} and \cite{AS:2010}, increases the power of moment inequality tests, while controlling size, by pre-selecting inequalities that are close to binding. In the specific implementation of moment inequality testing in \cite{HLS:2016}, the threshold according to which moment inequalities are pre-selected depends on the user-chosen quantities~$\kappa_n$ and~$B_n$.}. We choose the recommended values for the user-chosen parameters governing the generalized moment selection in \cite{HLS:2016}, namely $B_n=0.85\ln n/\ln\ln n$ and $\kappa_n=0.15\ln n$. To investigate robustness of the inference results to variations around this choice, we ran the tests in the case of the mother's education as an instrument for all pairs of values in $\{B_n/2,B_n,2B_n\}\times\{\kappa_n/2,\kappa_n,2\kappa_n\}$. Of the $48$ test results in the Canadian portions of Tables~\ref{table:IFm} and~\ref{table:Rm}, we see variation in the rejection level in one case only, related to Qu\'ebec.
In the German portions of Tables~\ref{table:IFm} and~\ref{table:Rm}, we see variation in rejection levels in four cases and reversal of the test results in three cases, related to East Germany.

Table~\ref{table:IFm} collects results of the test of the Roy model with imperfect foresight using the mother's education as an instrument satisfying Assumption~\ref{ass:SMIV} (SMIV).  The hypotheses that white men and women in the former Federal Republic choose their major to maximize expected income or the probability of a permanent employment a year after graduation are both rejected at the $1\%$ level. The hypothesis that white women in the former Democratic Republic choose their major to maximize the probability of a permanent employment a year after graduation is also rejected at the $10\%$ level. No other rejection of imperfect foresight Roy selection are found for residents of the former Democratic Republic or for minorities. 
 
The hypotheses that white women in the rest of Canada choose their major to maximize expected income or the probability of a permanent employment three years after graduation are both rejected at the $5\%$ level. The hypothesis that white women in Qu\'ebec choose their major to maximize the probability of securing employment related to their field of study is also rejected at the $5\%$ level. For men, Roy self-sorting is never rejected, which again shows a significantly different behavior for men and women. As in Germany, we find no rejections for visible minority men or women.
A notable feature of the results presented in Table~\ref{table:IFm} is that the hypothesis that white Canadian women's choices are driven by expected income or the probability of securing permanent employment is rejected for the rest of Canada, but not in Qu\'ebec, whereas the hypothesis that choices are driven by the probability of securing employment related to the field of study is rejected in Qu\'ebec, but not in the rest of Canada. This is consistent with the interpretation that labor market outcomes are stronger determinants of choices for women in Qu\'ebec, whereas field preferences are stronger determinants of choice for women in the rest of Canada.\footnote{\scriptsize To investigate this issue further, we tested a Roy model of self-sorting based on a variable equal to~1 when the applicant says they obtained the employment they were hoping for, and zero otherwise. This variable is available in Canada and the test result are identical to those obtained for the test of Roy self-sorting based on the relatedness of employment with field of specialization at university. Interpretation of this result, however, would hinge on a correct interpretation of the variable itself, which we do not have at this point.}

Table~\ref{table:Rm} reports results of the test of pure Roy self-sorting behavior based on three outcome variables, namely income, the degree to which employment is related to the field of study, and the vector (permanent, related) with lexicographically ordered components. We no longer include the ability to secure permanent employment, since it is a binary variable, and the tests of pure and imperfect foresight models are identical. As we see in Table~\ref{table:Rm}, the same conclusions hold for the pure Roy selection model, except that the hypothesis that white women in the former Democratic Republic choose their major to maximize expected income a year after graduation is now also rejected at the $1\%$ level, and the hypothesis that white women in Qu\'ebec choose their major to maximize the probability of securing permanent employment three years after graduation (and in case of ties decide based on relatedness of the employment) is now also rejected at the $10\%$ level. Again, there are no rejections of the pure Roy selection model for minorities anywhere, or for men anywhere in Canada.

Tables~\ref{table:IFmf} and~\ref{table:Rmf} collect similar results to those in Tables~\ref{table:IFm} and~\ref{table:Rm}, except that Assumption~\ref{ass:SMIV} (SMIV) holds for the vector of instruments combining mother's education and the proportion of women on the STEM faculty in the individual's region at the time of choice. Hence, only results for women are presented, since the proportion of women on the STEM faculty is conceived as a valid SMIV for women only. Again, there are no rejections of either the imperfect or the perfect foresight Roy models for minorities. 

The hypotheses that white women in the former Federal Republic choose their major to maximize expected income or the probability of a permanent employment a year after graduation are both rejected at the $1\%$ level.  The hypothesis that white women in the former Democratic Republic choose their major to maximize the their expected income (resp. probability of a permanent employment) a year after graduation is also rejected at the $1\%$ level. The same results hold for the test of perfect foresight Roy self-sorting.
Looking at the Canadian portion of Tables~\ref{table:IFmf} and~\ref{table:Rmf} reveals only slight discrepancies with test results with only the mother's education as the SMIV. 

For comparison, we look at older cohorts, based on the mother's education as an instrument (the proportion of women on the STEM being unavailable for these cohorts). We find rejection of Roy behavior based on income after one year (both perfect and imperfect foresight) for men from the former Federal Republic who obtained their degrees in $2005$, but not for women from the former Federal Republic or either gender from the former Democratic Republic. 
One initially surprising result in the Canadian portion of Table~\ref{table:IFmf} is the fact that Roy self-sorting
behavior for white women based on income is no longer rejected when the test is based on the
vector of instruments, whereas it was rejected based on mother’s education only. Although the theoretical bounds are tighter, an increased number of redundant moment inequalities reduces the power of the inference procedure.

Our results show a prevalence of rejections of Roy major selection behavior, possibly in favor of non pecuniary considerations, for categories that are generally considered privileged, particularly women, i.e., white women in Canada and white men and women from the former West German Federal Republic. We tend not to reject Roy major selection behavior for all other categories. This is borne out by the differences in responses to a survey question on the importance of labor market considerations on major choice. Table~\ref{table:LM} shows that minorities and residents of Qu\'ebec and the former Democratic Republic of Germany tend to weigh labor market considerations more than their counterparts.

To further investigate rejections of the Roy self-sorting behavior, we compute confidence intervals for the measure of departure from Roy (also called ``efficiency loss'') provided in Section~\ref{sec:dep} and~\ref{sec:depF}. We report the confidence lower bounds for white men and women from the former West Germany, for whom the Roy self-sorting behavior was rejected. For each of these categories, we plot the lower confidence bound as a function of income and the mother's education to identify regions of values (of income and mother's education) that are responsible for the rejection of Roy self-sorting. For white men in the former West Germany, we find that rejections are mostly driven by individuals, whose mothers earned postgraduate degrees. For white women in the former West Germany, we find rejections are driven by lower income women with high school educated mothers and median income women, whose mothers earned a high school degree only or a postgraduate degree.

Finally, we investigate the impact of Roy self-selection on income inequality in the case of individuals for whom the hypothesis of Roy self-sorting is not rejected, i.e., minorities of both genders in Germany and Canada, white women in the former East Germany and Canadian white men. Inference on the bounds from Propositions~\ref{prop:el} and~\ref{prop:el-gen} on the efficiency loss from non maximizing behavior and on the bounds (\ref{eq:low}) and (\ref{eq:iqr}) on the interquartile range in the randomized economy, is carried out with the STATA package {\em clrbounds} implementing \cite{CLR:2009}.
In Figure~\ref{fig:IQR}, we report confidence intervals for the partially identified interquartile range of potential non STEM income $Y_0$, potential STEM income $Y_1$, and aggregate income in an economy where individuals are randomized into sectors, next to the interquartile range for observed distributions of STEM, non STEM and aggregate income distributions. Most results are inconclusive, in the sense that realized interquartile ranges are well within the bounds for potential distributions, except in the case of white men in Qu\'ebec and white women in the former East Germany, where observed STEM interquartile range coincides with the lower bound on potential interquartile range
.




\section{Conclusion}

In this paper, we analyzed the Roy model of self-sorting into economic activities on the basis of anticipated outcomes. We stripped the model down to its essential features: we assumed that heterogeneous agents are characterized by a pair of potential outcomes, one for each sector of activity, and that they choose the sector that gives them a strictly higher outcome, leaving choice undetermined in case of ties. We characterized the restrictions this mechanism imposes on the joint distribution of potential outcomes. 
This characterization showed, on the one hand, that the Roy self-sorting mechanism puts non trivial restrictions on joint distributional features of potential outcomes, but, on the other hand, that the identified set is never empty, so that the Roy self-sorting mechanism described is not testable. Testability can be restored using selection shifters that are jointly independent of potential outcomes. However, such shifters are difficult to find in applications, and their usefulness is severely restricted by the Roy self-sorting mechanism, which only lets them affect selection in case potential outcomes are equal. We therefore introduced an extension of the notion of monotone instrumental variable, designed to constrain the joint distribution of potential outcomes, the quantity of interest. We considered (vectors of) variables that affect the vector of potential outcomes monotonically, in the sense of multivariate first order stochastic dominance, and called such (vectors of) variables stochastically monotone instrumental variables (SMIV). We repeated the characterization of the identified set for the joint distribution of potential outcomes under the SMIV assumption, and showed that testing the Roy self-sorting mechanism is equivalent to testing stochastic monotonicity of observed outcomes in the instrument. To alleviate the concern that rejections are due to the assumption that agents are perfectly informed of their future outcomes, we repeated the exercise with an imperfect foresight version of the model, where agents select sectors based on expectations. Beyond testing the Roy self-sorting mechanism and providing measures of departure from outcome-based decisions, we highlighted another important application of our characterization of the identified set of joint potential outcome distributions, namely the derivation of sharp bounds on the interquantile range of potential outcome distributions to revisit the effects of self selection on inequality in employment outcomes.

We applied our methodology to the analysis of major choices made by graduates of Canadian and German universities based on the national graduate surveys of each of these two countries. We analyzed selection of mathematics-intensive versus other fields of study by graduates within the framework of the Roy model with employment based outcomes that include income 1 and 3 years after graduation, the ability to secure permanent employment by the time of the survey and the extent to which employment secured is related to the field of study. The data supports previous evidence of a labor market advantage of mathematics-intensive fields (STEM), severe under-representation of women in STEM, over-representation of visible minorities in STEM and male labor market advantage in both sectors. We investigated whether selection behavior is consistent with Roy self-sorting on outcomes for categories of graduates by gender, visible minority status and region of residence (former East and former West Germany, Qu\'ebec and the rest of Canada). To test Roy self-sorting based on employment outcomes, we used parental education level and the proportion of women on the faculty of STEM programs in the region and at the time of choice as stochastically monotone instruments. We found a pattern of rejections of Roy self-sorting based on outcomes for white women in the former Federal Republic of Germany and the rest of Canada, and a lack of rejections for visible minorities and for white women from Qu\'ebec and white men from all of Canada and the former German Democratic Republic. Confidence intervals for measures of departure from Roy behavior revealed that in the case of white women from the former Federal Republic, for instance, rejection of Roy behavior seems to be driven by lower income women with high school educated mothers and middle income women with postgraduate educated mothers. Among groups, where Roy self-sorting is not rejected, comparisons of interquartile ranges for observed and counterfactual income distributions are inconclusive except in the cases of women in the former Democratic Republic and white men from Qu\'ebec, where self-sorting is found not to increase inequality.

The pattern of rejections of Roy self-sorting in major choice points to non labor market related determinants of choice. For instance, our results are consistent with a story involving gender profiling pushing white men in the West of Germany into STEM fields and white women in the West of Germany and in Canada out of STEM fields. They are also consistent with gender profiling being less prevalent in the former communist Germany. However, the results are also consistent with a story involving non pecuniary field preferences driving major choices of more privileged groups in more affluent regions, but not the choices of the more financially constrained. The methodology proposed here should then be construed as a tool for exploratory analysis of the determinants of major choice prior to a fully structural generalized Roy modeling of preferences, \`a la \cite{HV:99}, \cite{HV:2005}, in a context where, unlike the analysis of returns to college, there is no clear cost differential between different choices. Non rejections of Roy self-selection based on labor market outcomes, on the other hand, are a warning that policies directly aimed at increasing the share of women in STEM majors at university may have a (possibly short term) negative effect on the gender gap and wage inequality, and that both upstream (early childhood) and downstream (labor market) interventions are required.




\begin{appendix}

{\scriptsize

\section{Proofs and additional results relating to binary outcomes}
\label{app:proofs}

\subsection{Sharp bounds for the binary outcome Roy model}

\subsubsection*{Statement of Proposition~\ref{prop:sharp}}
Fix the pair of binary random variables $(Y,D)$ with probability mass function $(q_{00},q_{01},q_{10},q_{11})$, with $q_{ij}:=\mathbb P(Y=i,D=j)$. The following two statements hold. (1) If the non negative vector $(p_{00},p_{01},p_{10},p_{11})\in\mathbb R^4$ satisfies $p_{00}+p_{01}+p_{10}+p_{11}=1$,
$p_{10}\leq q_{10}$, $p_{01}\leq q_{11}$ and $p_{00}=q_{00}+q_{01}$, then there exists a pair of binary random variables $(Y_0,Y_1)$ such that Assumptions~\ref{ass:roy} and~\ref{ass:sel} are satisfied and $\mathbb P(Y_0=0,Y_1=0)=p_{00}$, $\mathbb P(Y_0=0,Y_1=1)=p_{01}$, $\mathbb P(Y_0=1,Y_1=0)=p_{10}$ and $\mathbb P(Y_0=1,Y_1=1)=p_{11}$.
(2) Conversely, if the pair of binary random variables $(Y_0,Y_1)$ satisfies Assumptions~\ref{ass:roy} and~\ref{ass:sel}, then $\mathbb P(Y_0=1,Y_1=0)\leq q_{10}$, $\mathbb P(Y_0=0,Y_1=1)\leq q_{11}$ and $\mathbb P(Y_0=0,Y_1=0)=q_{00}+q_{01}$.

\subsubsection*{Proof of Proposition~\ref{prop:sharp}}
Write $p_{ij}:=\mathbb P(Y_0=i,Y_1=j)$ for each $i,j=0,1$. The binary outcomes Roy model of Definition~\ref{def:bin} can be equivalently defined as a correspondence $G$ between 
values of observables $(y,d)\in\mathcal A:=\{(0,0),(0,1),(1,0),(1,1)\}$ and values of unobservables $(y_0,y_1)\in\mathcal A$.
The correspondence is defined by its values $G(y,d)$ for each $(y,d)\in\mathcal A$, namely $G(1,1):=\{(1,1),(0,1)\}$, $G(1,0):=\{(1,1),(1,0)\}$, $G(0,1):=\{(0,0)\}$ and $G(0,0):=\{(0,0)\}$.
By Theorem~1 of \cite{GH:2011}, the $14$ inequalities $\mathbb P((Y_0,Y_1)\in A)\leq \mathbb P(G(Y,D)\cap A\ne\varnothing)$ for each $A\subset\mathcal A$ provide a collection of sharp bounds for the model defined by the correspondence $G$. For instance, $A=\{(0,0)\}$ yields the inequality $p_{00}\leq q_{00}+q_{01}$ and $A=\{(1,1),(0,1)\}$ yields the inequality $p_{11}+p_{01}\leq q_{11}+q_{10}$. To prove the result, it suffices to show that all $14$ inequalities are implied by $0\leq p_{10}\leq q_{10}$, $0\leq p_{01}\leq q_{11}$ and $p_{00}=q_{00}+q_{01}$. The $14$ inequalities are listed below.
Singleton $A$'s yield
\begin{eqnarray}
\label{sing}
\begin{array}{lll}
p_{11}&\leq& q_{11}+q_{10}\\
p_{10}&\leq& q_{10}\\
p_{01}&\leq& q_{11}\\
p_{00}&\leq& q_{01}+q_{00}.
\end{array}
\end{eqnarray}
Pairs yield
\begin{eqnarray}
\label{pai}
\begin{array}{lll}
p_{11}+p_{10}&\leq& q_{11}+q_{10}\\
p_{11}+p_{01}&\leq& q_{11}+q_{10}\\
p_{11}+p_{00}&\leq& 1\\
p_{10}+p_{01}&\leq& q_{11}+q_{10}\\
p_{10}+p_{00}&\leq& q_{10}+q_{01}+q_{00}\\
p_{01}+p_{00}&\leq& q_{11}+q_{01}+q_{00}.
\end{array}
\end{eqnarray}
Finally, triplets yield
\begin{eqnarray}
\label{tri}
\begin{array}{lll}
p_{11}+p_{10}+p_{01}&\leq& q_{11}+q_{10}\\
p_{11}+p_{10}+p_{00}&\leq& 1\\
p_{11}+p_{01}+p_{00}&\leq& 1\\
p_{10}+p_{01}+p_{00}&\leq& 1.
\end{array}
\end{eqnarray}
The first four inequalities in (\ref{pai}) are implied by the first inequality in (\ref{tri}). The last two are implied by (\ref{sing}). All inequalities in (\ref{pai}) are therefore redundant. Since $p_{11}=1-p_{00}-p_{01}-p_{10}$, all four inequalities in (\ref{sing}) are implied by $0\leq p_{10}\leq q_{10}$, $0\leq p_{01}\leq q_{11}$ and $p_{00}=q_{00}+q_{01}$. Finally, since $p_{11}+p_{10}+p_{01}=1-p_{00}$, the first inequality in (\ref{tri}) is implied by $p_{00}=q_{00}+q_{01}$ and the result follows.

\subsubsection*{Extension to the {\em alternative binary Roy model}}
(1) First, we show that $\mathbb P(Y_0=1,Y_1=0)\leq q_{10}$, $\mathbb P(Y_0=0,Y_1=1)\leq q_{11}$ and $\mathbb P(Y_0=0,Y_1=0)=q_{00}+q_{01}$ hold if $(Y,D,Y_0,Y_1)$ satisfy the assumptions of Definition~\ref{def:alt}. Under the specification of Definition~\ref{def:alt}, $Y_0=1$ and $Y_1=0$ jointly imply that $Y_0^\ast>Y_1^\ast$, which in turn implies $D=0$ and $Y=1$, so that the first inequality holds. The second holds by the same reasoning and the roles of $Y_0$ and $Y_1$ reversed. Finally, $Y_0=Y_1=0$ implies $Y=0$, and $Y=1$ implies that $Y_0=1$ or $Y_1=1$, so the equality holds as well. 
(2) Second, the binary outcomes Roy model specification of Definition~\ref{def:bin} is nested in the alternative binary Roy model specification of Definition~\ref{def:alt}. Indeed, the former can be obtained by restricting $(Y_0^\ast,Y_1^\ast)$ to be binary. Hence, sharpness of the bounds for the binary outcomes Roy model implies sharpness for the alternative binary Roy model. The result follows.

\subsubsection*{Representation of the bounds on the $2$-simplex}
We continue to denote $\mathbb P(Y=i,D=j)=q_{ij}$ and $\mathbb P(Y_0=i,Y_1=j)=p_{ij}$. According to Proposition~\ref{prop:sharp}, $p_{00}=q_{01}+q_{00}$. Hence, the remaining three probabilities, namely $p_{10}$, $p_{01}$ and $p_{11}=q_{11}+q_{10}-p_{10}-p_{01}$ can be represented in barycentric coordinates in the rescaled $2$-simplex of Figure~\ref{fig:2simplex}, where the three vertices correspond to the cases, where $p_{11}=q_{11}+q_{10}$, $p_{10}=q_{11}+q_{10}$ and $p_{01}=q_{11}+q_{10}$ respectively.

\begin{figure}[htbp]
\begin{center}
\caption{{\scriptsize Identified set for $(p_{10},p_{01},p_{11}=q_{11}+q_{10}-p_{10}-p_{01})$ in barycentric coordinates in the rescaled $2$-simplex. $p_{00}$ is identified and equal to $q_{01}+q_{00}$. The left-hand-side figure is without excluded variable $Z$. The right-hand-side is in the presence of variation in a variable~$Z$ satisfying Assumption~\ref{ass:SMIV}. The conditioning variable $z$ is omitted from the notation in the graph and $\bar{q}_{1j}(z):=\sup_{\tilde z\leq z}q_{1j}(\tilde z)$, $\underline{q}_{1j}(z):=q_{11}(z)+q_{10}(z)-\bar{q}_{1,1-j}(z)$,~$j=0,1$.}}
\label{fig:2simplex}
\vskip15pt

\includegraphics[width=0.9\textwidth]{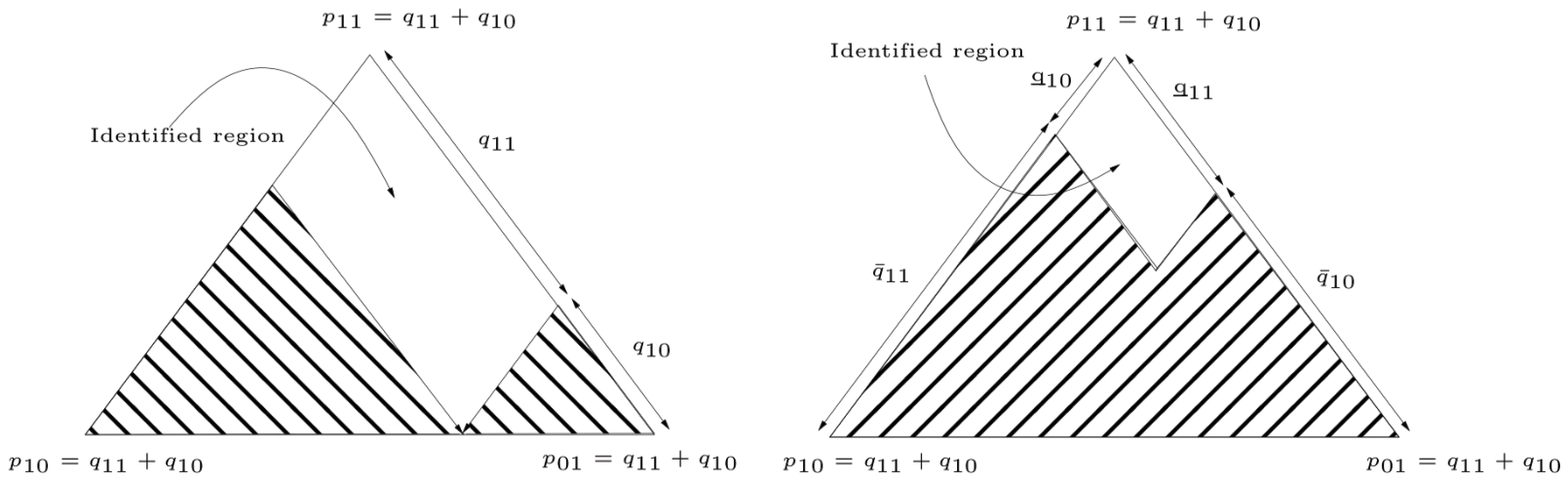}

\end{center}
\end{figure}


\subsection{Covariate restrictions}\label{app:cov}
\subsubsection*{Statement of Theorem~\ref{thm:SMIV}(1)}
Fix the joint distribution $(Y,D,Z)$ and denote the conditional probability mass function $(q_{00}(z),q_{01}(z),q_{10}(z),q_{11}(z))$, where $q_{ij}(z)=\mathbb P(Y=i,D=j\vert Z=z)$. The following two statements hold. (1) If for each $z\in$ Supp$(Z)$, the non-negative vector $(p_{00}(z),p_{01}(z),p_{10}(z),p_{11}(z))\in\mathbb R^4$ satisfies
$p_{00}(z)+p_{01}(z)+p_{10}(z)+p_{11}(z)=1$, $\sup_{\tilde z\leq z}q_{10}(\tilde z)\leq p_{01}(z)+p_{11}(z)$, $\sup_{\tilde z\leq z}q_{11}(\tilde z)\leq p_{10}(z)+p_{11}(z)$ and $p_{00}(z)=q_{01}(z)+q_{00}(z)$, then there exists a pair of binary random variables $(Y_0,Y_1)$ such that Assumptions~\ref{ass:roy}, ~\ref{ass:sel} and~\ref{ass:SMIV} are satisfied and $\mathbb P(Y_0=0,Y_1=0\vert Z=z)=p_{00}(z)$, $\mathbb P(Y_0=0,Y_1=1\vert Z=z)=p_{01}(z)$ and $\mathbb P(Y_0=1,Y_1=0\vert Z=z)=p_{10}(z)$, for each $z\in$ Supp$(Z)$.
(2) Conversely, if the random vector $(Y,D,Z)$ satisfies Assumptions~\ref{ass:roy}, \ref{ass:sel} and~\ref{ass:SMIV} for some the pair of binary random variables $(Y_0,Y_1)$, then~(\ref{eq:SMIV}) holds.

\subsubsection*{Proof of Theorem~\ref{thm:SMIV}(1)}
From the proof of Proposition~\ref{prop:sharp}, the identified set under Assumptions~\ref{ass:roy} and~\ref{ass:sel} is characterized by $q_{10}(z)\leq p_{10}(z)+p_{11}(z)\leq q_{10}(z)+q_{11}(z)$, $q_{11}(z)\leq p_{01}(z)+p_{11}(z)\leq q_{10}(z)+q_{11}(z)$ and $p_{00}(z)=q_{01}(z)+q_{00}(z)$ for all $z\in$ Supp$(Z)$. Assumption~\ref{ass:SMIV} is equivalent to $\mathbb P((Y_0,Y_1)\in U\vert Z=z_1)\leq\mathbb P((Y_0,Y_1)\in U\vert Z=z_2)$ for all $z_1\leq z_2$ and all upper set $U$ in $\{0,1\}^2$. The upper sets are $\{(1,1)\}, \{(1,1),(1,0)\}$, \{(1,1),(0,1)\} and $\{(1,1),(1,0),(0,1)\}$. Hence, Assumption~\ref{ass:SMIV} is equivalent to $\sup_{\tilde z\leq z}p_{11}(\tilde z)\leq p_{11}(z)\leq\inf_{\tilde z\geq z}p_{11}(\tilde z)$, $\sup_{\tilde z\leq z}[p_{11}(\tilde z)+p_{10}(\tilde z)]\leq p_{11}(z)+p_{10}(z)\leq\inf_{\tilde z\geq z}[p_{11}(\tilde z)+p_{10}(\tilde z)]$, $\sup_{\tilde z\leq z}[p_{11}(\tilde z)+p_{01}(\tilde z)]\leq p_{11}(z)+p_{01}(z)\leq\inf_{\tilde z\geq z}[p_{11}(\tilde z)+p_{01}(\tilde z)]$, and $\sup_{\tilde z\leq z}[1-p_{00}(\tilde z)]\leq 1-p_{00}(z)\leq\inf_{\tilde z\geq z}[1-p_{00}(\tilde z)]$ for all $z\in$ Supp$(Z)$. Combining the two sets of inequalities yields the result.

\subsubsection*{Sector specific exclusions}
We denote by $X_d$ the vector of observable variables (if any) that enter in the equation determining potential outcome $Y_d$, but not $Y_{1-d}$. Since there is some ambiguity in notation, it is worth stressing the fact that both vectors $X_0$ and $X_1$ are observed, irrespective of the chosen sector, unlike $Y_0$, which is only observed if $D=0$ and $Y_1$, when $D=1$.
\begin{assumption}[Sector specific exclusions]
\label{ass:ex}
The random vectors $X_0$ and $X_1$ denote vectors of observed variables (when they exist) such that $Y_d\perp\!\!\!\perp X_{1-d}|X_d$, for $d=0$ and $1.$
\end{assumption}
The exclusions of Assumption~\ref{ass:ex} are conditional on a set of additional observed covariates, as noted before.
Excluded variables $X_d$ are variables that change the price of skills relevant for one sector without affecting the price of skills in the other, as discussed in \cite{HH:90}. Typical examples would include sector specific shifters of labor market conditions, as in \cite{HS:85}, \cite{HS:90}. In the case of college major choice, considered in Section~\ref{sec:appli}, in a narrow partial equilibrium sense, exogenous and unanticipated variation (at the time of college major choice) in the gross number of STEM jobs could be thought to affect only conditions for success in securing employment with a STEM degree, without affecting success in securing employment with a non STEM degree.

The classical way to derive bounds under an exclusion restriction is to observe that $\mathbb E(Y_d\vert X_d,X_{1-d})=\mathbb E(Y_d\vert X_d)$ under Assumption~\ref{ass:ex}, so that the bounds (\ref{eq:marg}) hold for all values of $X_{1-d}$. We contribute to the literature here, in showing sharpness of these bounds for the binary (and alternative binary) Roy model. Conditioning on all non excluded variables remains implicit throughout. 
\begin{proposition}[Marginal bounds with sector specific covariates]
\label{prop:ex}
For any $(x_0,x_1)$ in the support of $(X_0,X_1)$, the identified set for the parameter vector $(\mathbb E(Y_0\vert X_0=x_0),\mathbb E(Y_1\vert X_1=x_1))$ in the binary (and alternative binary) Roy model is characterized by:
\begin{eqnarray*}
\begin{array}{lclcl}
\mathbb P(Y=1,D=0\vert X_0=x_0,X_1=\tilde x_1)&\leq&\mathbb E(Y_0\vert X_0=x_0)&\leq&\mathbb  P(Y=1\vert  X_0=x_0,X_1=\tilde x_1),\\\\
\mathbb P(Y=1,D=1\vert X_0=\tilde x_0, X_1=x_1)&\leq&\mathbb E(Y_1\vert X_1=x_1)&\leq&\mathbb  P(Y=1\vert X_0=\tilde x_0, X_1=x_1),
\end{array}
\end{eqnarray*}
for almost all $\tilde x_1\in$ Supp$(X_1\vert X_0=x_0)$, and $\tilde x_0\in$ Supp$(X_0\vert X_1=x_1)$.
\end{proposition}
The bounds define the identified set for the vector $(\mathbb E(Y_0\vert X_0=x_0),\mathbb E(Y_1\vert X_1=x_1))$, namely, any value of that vector satisfying the bounds can be achieved as a solution of the model for some distribution
of the observable variables $(Y,D)$ conditional on $(X_0=x_0,X_1=x_1)$. In other words, no value for the pair $(\mathbb E(Y_0\vert X_0=x_0),\mathbb E(Y_1\vert X_1=x_1))$ that satisfies both equations can be rejected solely on the basis of the model specification. The bounds are well-known, but the joint sharpness result is new. As before, the bounds of Proposition~\ref{prop:ex} are {\em intersection bounds}, so that inference can be carried out with the method proposed in \cite{CLR:2009}.

A salient consequence of Proposition~\ref{prop:ex} is the fact that the binary outcomes Roy model can be rejected when the bounds cross, i.e., when there is a value $x_1$ in the support of $X_1$ and two values $x_0^1$ and $x_0^2$ in the support of $X_0$ conditional on $X_1=x_1$, such that $\mathbb P(Y=1,D=1\vert X_0=x_0^1,X_1=x_1)>\mathbb P(Y=1\vert X_0=x_0^2,X_1=x_1)$ or a value $x_0$ in the support of $X_0$ and two values $x_1^1$ and $x_1^2$ in the support of $X_1$ conditional on $X_0=x_0$, such that $\mathbb P(Y=1,D=0\vert X_0=x_0,X_1=x_1^1)>\mathbb P(Y=1\vert X_0=x_0,X_1=x_1^2)$. 
Identification of the pair $(\mathbb E(Y_0\vert X_0=x_0),\mathbb E(Y_1\vert X_1=x_1))$ can be achieved as a simple implication of the previous result if there is $\tilde x_1\in$ Supp$(X_1\vert X_0=x_0)$ such that $\mathbb P(Y=1,D=1\vert X_0=x_0,X_1=\tilde x_1)=0$ and $\tilde x_0\in$ Supp$(X_0\vert X_1=x_1)$ such that $\mathbb P(Y=1,D=0\vert X_0=\tilde x_0,X_1=x_1)=0$, in which case lower and upper bounds coincide in Proposition~\ref{prop:ex}. This identification result is akin to the {\em identification at infinity} of \cite{Heckman:90}.

\subsubsection*{Proof of Proposition~\ref{prop:ex}}
Validity of the bounds was shown above. For sharpness, fix $(x_0,x_1)$ in the Support of $(X_0,X_1)$. For a given random vector $(Y,D)$ of binary random variables, denote by $q_{ij}(\tilde x_0,\tilde x_1)$ the conditional probability $\mathbb P(Y=i,D=j\vert X_0=\tilde x_0,X_1=\tilde x_1)$ for any $(\tilde x_0,\tilde x_1)$ in the Support of $(X_0,X_1)$. Consider any pair $(a(x_0),b(x_1))$ satisfying 
\begin{eqnarray}
\label{eq:a}
q_{10}(x_0,\tilde x_1)\leq a(x_0)\leq q_{11}(x_0,\tilde x_1)+q_{10}(x_0,\tilde x_1)
\end{eqnarray}
for almost all $\tilde x_1\in$ Supp$(X_1\vert X_0=x_0)$, and 
\begin{eqnarray}
\label{eq:b}
q_{11}(\tilde x_0,x_1)\leq b(x_1)\leq q_{11}(\tilde x_0,x_1)+q_{10}(\tilde x_0,x_1)
\end{eqnarray} 
for almost all $\tilde x_0\in$ Supp$(X_0\vert X_1=x_1)$. We exhibit a pair of binary random variables $(Y_0,Y_1)$ with joint distribution $p_{ij}:=\mathbb P(Y_0=i,Y_1=j\vert X_0=x_0,X_1=x_1)$, such that Assumptions~\ref{ass:roy}, \ref{ass:sel} and~\ref{ass:ex} are satisfied, and such that
\begin{eqnarray}
\label{eq:c}
p_{11}(x_0,x_1)+p_{10}(x_0,x_1)=a(x_0) \mbox{ and } p_{11}(x_0,x_1)+p_{01}(x_0,x_1)=b(x_1).
\end{eqnarray}
Here is our proposed distribution.
\begin{eqnarray*}
p_{00}(x_0,x_1)&=&q_{00}(x_0,x_1) + q_{01}(x_0,x_1), \label{eqp00}\\
p_{11}(x_0,x_1)&=&b(x_1) + a(x_0)-q_{10}(x_0,x_1) - q_{11}(x_0,x_1),\label{eqp11}\\
p_{10}(x_0,x_1)&=&q_{10}(x_0,x_1) +q_{11}(x_0,x_1)-b(x_1),\label{eqp10}\\
p_{01}(x_0,x_1)&=&q_{10}(x_0,x_1) +q_{11}(x_0,x_1)-a(x_0).\label{eqp01}
\end{eqnarray*}
Note that (\ref{eq:c}) is verified by construction. We also verify that $p_{00}(x_0,x_1)+p_{01}(x_0,x_1)+p_{10}(x_0,x_1)+p_{11}(x_0,x_1)=1$ and that 
$p_{00}$, $p_{10}$,  and $p_{01}$ are nonnegative.
From (\ref{eq:a}) and (\ref{eq:b}), $q_{10}(x_1,x_0) + q_{11}(x_1,x_0)  \leq  a(x_0) + b(x_1)$,  which implies that $p_{11}(x_0,x_1)$ is also nonnegative. 
Assumption~\ref{ass:ex} is implied by (\ref{eq:c}) irrespective of the construction of $(Y_0,Y_1)$. We now construct a pair $(Y_0,Y_1)$ with conditional distribution $p_{ij}(x_0,x_1)$ such that Assumptions~\ref{ass:roy} and~\ref{ass:sel} are both satisfied. First construct a random variable $U$ with uniform distribution on $[0,1]$ in the following way. Set $U\in[0,q_{00}(x_0,x_1)+q_{01}(x_0,x_1)]$ if and only if $Y=0$. Set $U\in(q_{00}(x_0,x_1)+q_{01}(x_0,x_1),q_{00}(x_0,x_1)+q_{01}(x_0,x_1)+q_{10}(x_0,x_1)]$ if and only if $(Y,D)=(1,0)$. Finally, set $U\in(q_{00}(x_0,x_1)+q_{01}(x_0,x_1)+q_{10}(x_0,x_1),1]$ if and only if $(Y,D)=(1,1)$. 
Now set $(Y_0,Y_1)=(0,0)$ if and only if $U\leq q_{00}(x_0,x_1)+q_{01}(x_0,x_1)$, $(Y_0,Y_1)=(1,0)$ if and only if $U\in(q_{00}(x_0,x_1)+q_{01}(x_0,x_1),q_{00}(x_0,x_1)+q_{01}(x_0,x_1)+p_{10}(x_0,x_1)]$, $(Y_0,Y_1)=(1,1)$ if and only if $U\in(q_{00}(x_0,x_1)+q_{01}(x_0,x_1)+p_{10}(x_0,x_1),1-p_{01}(x_0,x_1)]$, and $(Y_0,Y_1)=(0,1)$ if and only if $U\in(1-p_{01}(x_0,x_1),1]$. By construction, $(Y_0,Y_1)$ has probability mass distribution $p_{ij}(x_0,x_1)$ and satisfies Assumptions~\ref{ass:roy} and~\ref{ass:sel}. This completes the proof.

\subsubsection*{Statement of Theorem~\ref{thm:if}(1)}
Fix the joint distribution $(Y,D,Z)$ and denote the conditional probability mass function $(q_{00}(z),q_{01}(z),q_{10}(z),q_{11}(z))$, where $q_{ij}(z)=\mathbb P(Y=i,D=j\vert Z=z)$. The following two statements hold. (1) If for each $z\in$ Supp$(Z)$, the non-negative vector $(p_{00}(z),p_{01}(z),p_{10}(z),p_{11}(z))\in\mathbb R^4$ satisfies
$p_{00}(z)+p_{01}(z)+p_{10}(z)+p_{11}(z)=1$, $\sup_{\tilde z\leq z}q_{10}(\tilde z)\leq p_{10}(z)+p_{11}(z)\leq\inf_{\tilde z\geq z}[q_{11}(\tilde z)+q_{10}(\tilde z)]$, $\sup_{\tilde z\leq z}q_{11}(\tilde z)\leq p_{01}(z)+p_{11}(z)\leq\inf_{\tilde z\geq z}[q_{11}(\tilde z)+q_{10}(\tilde z)]$, $p_{00}(z)\leq\inf_{\tilde z\leq z}[q_{01}(\tilde z)+q_{00}(\tilde z)]$, $p_{10}(z)\leq q_{01}(z)+q_{10}(z)$, and $p_{01}(z)\leq q_{00}(z)+q_{11}(z)$, then there exists a pair of binary random variables $(Y_0,Y_1)$ such that Assumptions~\ref{ass:roy}, ~\ref{ass:if} and~\ref{ass:SMIV} are satisfied and $\mathbb P(Y_0=0,Y_1=0\vert Z=z)=p_{00}(z)$, $\mathbb P(Y_0=0,Y_1=1\vert Z=z)=p_{01}(z)$ and $\mathbb P(Y_0=1,Y_1=0\vert Z=z)=p_{10}(z)$, for each $z\in$ Supp$(Z)$.
(2) Conversely, if the random vector $(Y,D,Z)$ satisfies Assumptions~\ref{ass:roy}, \ref{ass:if} and~\ref{ass:SMIV} for some the pair of binary random variables $(Y_0,Y_1)$, then~(\ref{eq:gen-if}) and~(\ref{eq:marg-if})~hold.

\subsubsection*{Statement of Proposition~\ref{prop:el}(1)}
Fix the joint distribution $(Y,D,Z)$ and denote the conditional probability mass function $(q_{00}(z),q_{01}(z),q_{10}(z),q_{11}(z))$, where $q_{ij}(z)=\mathbb P(Y=i,D=j\vert Z=z)$. The following two statements hold. (1) If for each $z\in$ Supp$(Z)$, the non-negative vector $(p_{00}(z),p_{01}(z),p_{10}(z),p_{11}(z))\in\mathbb R^4$ satisfies
$p_{00}(z)+p_{01}(z)+p_{10}(z)+p_{11}(z)=1$, $\sup_{\tilde z\leq z}q_{10}(\tilde z)\leq p_{10}(z)+p_{11}(z)\leq1-\sup_{\tilde z\geq z}q_{00}(\tilde z)$, $\sup_{\tilde z\leq z}q_{11}(\tilde z)\leq p_{01}(z)+p_{11}(z)\leq1-\sup_{\tilde z\geq z}q_{01}(\tilde z)$, $p_{11}(z)\leq\inf_{\tilde z\geq z}[q_{10}(\tilde z)+q_{11}(\tilde z)]$, $p_{00}(z)\leq\inf_{\tilde z\leq z}[q_{01}(\tilde z)+q_{00}(\tilde z)]$, $p_{10}(z)\leq q_{01}(z)+q_{10}(z)$, and $p_{01}(z)\leq q_{00}(z)+q_{11}(z)$, then there exists a pair of binary random variables $(Y_0,Y_1)$ such that Assumptions~\ref{ass:roy} and~\ref{ass:SMIV} are satisfied and $\mathbb P(Y_0=0,Y_1=0\vert Z=z)=p_{00}(z)$, $\mathbb P(Y_0=0,Y_1=1\vert Z=z)=p_{01}(z)$ and $\mathbb P(Y_0=1,Y_1=0\vert Z=z)=p_{10}(z)$, for each $z\in$ Supp$(Z)$.
(2) Conversely, if the random vector $(Y,D,Z)$ satisfies Assumptions~\ref{ass:roy} and~\ref{ass:SMIV} for some the pair of binary random variables $(Y_0,Y_1)$, then~(\ref{eq:gen-smiv}) and~(\ref{eq:marg-smiv})~hold.

\subsubsection*{Proof of Proposition~\ref{prop:el}(1)}
For each $z$ in the support of $Z$, the binary outcomes model under Assumption~\ref{ass:roy} can be equivalently defined as a correspondence $G$ between 
values of observables $(y,d)\in\mathcal A:=\{(0,0),(0,1),(1,0),(1,1)\}$ and values of unobservables $(y_0,y_1)\in\mathcal A$.
The correspondence is defined by its values $G(y,d)$ for each $(y,d)\in\mathcal A$, namely $G(1,1):=\{(1,1),(0,1)\}$, $G(1,0):=\{(1,1),(1,0)\}$, $G(0,1):=\{(1,0),(0,0)\}$ and $G(0,0):=\{(0,1),(0,0)\}$.
By Theorem~1 of \cite{GH:2011}, the $14$ inequalities $\mathbb P((Y_0,Y_1)\in A\vert Z=z)\leq \mathbb P(G(Y,D)\cap A\ne\varnothing\vert Z=z)$ for each $A\subset\mathcal A$ provide a collection of sharp bounds for the model defined by the correspondence $G$. The $14$ inequalities are listed below.
Singleton $A$'s yield
\begin{eqnarray}
\label{single}
\begin{array}{lll}
p_{11}(z)&\leq& q_{11}(z)+q_{10}(z)\\
p_{10}(z)&\leq& q_{10}(z)+q_{01}(z)\\
p_{01}(z)&\leq& q_{11}(z)+q_{00}(z)\\
p_{00}(z)&\leq& q_{01}(z)+q_{00}(z).
\end{array}
\end{eqnarray}
Pairs yield
\begin{eqnarray}
\label{pairs}
\begin{array}{lll}
p_{11}(z)+p_{10}(z)&\leq& q_{11}(z)+q_{10}(z)+q_{01}(z)\\
p_{11}(z)+p_{01}(z)&\leq& q_{11}(z)+q_{10}(z)+q_{00}(z)\\
p_{11}(z)+p_{00}(z)&\leq& 1\\
p_{10}(z)+p_{01}(z)&\leq& 1\\
p_{10}(z)+p_{00}(z)&\leq& q_{10}(z)+q_{01}(z)+q_{00}(z)\\
p_{01}(z)+p_{00}(z)&\leq& q_{11}(z)+q_{01}(z)+q_{00}(z).
\end{array}
\end{eqnarray}
Finally, triplets yield only trivial inequalities of the form $p_{11}(z)+p_{10}(z)+p_{01}(z)\leq 1$.

All non trivial inequalities in (\ref{pairs}) are equivalent to $q_{10}(z)\leq p_{10}(z)+p_{11}(z)\leq 1-q_{00}(z)$ and  $q_{11}(z)\leq p_{01}(z)+p_{11}(z)\leq 1-q_{01}(z)$.  Combining with Assumption~\ref{ass:SMIV} as in the proof of Theorem~\ref{thm:SMIV}(1) and removing redundant inequalities yields the result.

\subsubsection*{Proof of Theorem~\ref{thm:if}(1)}
Assumption~\ref{ass:if} is equivalent to $Y=Y_d\Rightarrow\mathbb E[Y\vert \mathcal I]=\mathbb E[Y\vert \mathcal I]\geq\mathbb E[Y_{1-d}\vert \mathcal I]$ for $d=0,1.$ The latter statement is true for some $\sigma$-algebra that contains $\sigma(Z)$ if and only if $Y=Y_d\Rightarrow\mathbb E[Y\vert Z]=\mathbb E[Y_d\vert Z]\geq\mathbb E[Y_{1-d}\vert Z]$ for $d=0,1.$ The latter is equivalent to $\max\{p_{01}(z)+p_{11}(z),p_{10}(z)+p_{11}(z)\}\leq q_{11}(z)+q_{10}(z)$ for all $z\in$ Supp$(Z)$. Combining with~(\ref{single}) and~(\ref{pairs}) and removing redundant inequalities, yields $q_{10}(z)\leq p_{01}(z)+p_{11}(z)\leq q_{11}(z)+q_{10}(z)$, $q_{11}(z)\leq p_{10}(z)+p_{11}(z)\leq q_{11}(z)+q_{10}(z)$, $p_{00}(z)\leq q_{01}(z)+q_{00}(z)$, $p_{10}(z)\leq q_{01}(z)+q_{10}(z)$, and $p_{01}(z)\leq q_{00}(z)+q_{11}(z)$ for all $z\in$ Supp$(Z)$. Combining with Assumption~\ref{ass:SMIV} as in the proof of Theorem~\ref{thm:SMIV}(1) yields the result.

\subsubsection*{Combined sector-specific and SMIV instrument}
Suppose $Z$ satisfies Assumption~\ref{ass:SMIV} (SMIV) and $Y_0\perp\!\!\!\perp Z$, so that $Z$ is both a stochastically monotone instrument and a sector specific variable in the sense that it does not directly affect potential outcomes in the non STEM sector. Then, the joint distribution of potential outcomes in the binary outcomes Roy model (Assumptions~\ref{ass:roy} and~\ref{ass:sel}) satisfies
{ \scriptsize
\begin{eqnarray*}
\begin{array}{ccl}
\sup_{z}\mathbb P(Y=1,D=0\vert Z=z)&\leq&\mathbb P(Y_0=1)\;\;\leq\;\;\inf_{z}\mathbb P(Y=1\vert Z=z)\\\\
\sup_{\tilde z\leq z}\mathbb P(Y=1,D=1\vert Z=\tilde z)&\leq&\mathbb P(Y_1=1\vert Z=z)\\\\
\mathbb P(Y_0=Y_1=0\vert Z=z)&=&\mathbb P(Y=0\vert Z=z).
\end{array}
\end{eqnarray*}
}
Testable implications are stochastic monotonicity of $Y$ relative to $Z$ and 
{\scriptsize 
\[
\sup_{z}\mathbb P(Y=1,D=0\vert Z=z)\leq\inf_{z}\mathbb P(Y=1\vert Z=z).
\]
}


\section{Proofs and additional results relating to mixed discrete-continuous outcomes}

\subsection{Functionally sharp bounds for the Roy model}
\label{sec:pjoint}
We first illustrate functional sharpness by showing improvements over Peterson bounds. Combining Peterson bounds (\ref{eq:pjoint}) and assuming $y_{12}>y_{01}$ and $y_{02}>y_{11}$ yields the following upper bound.
\begin{eqnarray*}
\mathbb P(y_{01}<Y_0\leq y_{02},y_{11}<Y_1\leq y_{12})&=&
\mathbb P(Y_0\leq y_{02},Y_1\leq y_{12})-\mathbb P(Y_0\leq y_{02},Y_1\leq y_{11})\\
&&-\mathbb P(Y_0\leq y_{01},Y_1\leq y_{12})+\mathbb P(Y_0\leq y_{01},Y_1\leq y_{11})\\
&\leq&\mathbb P(Y\leq y_{02},D=0)+\mathbb P(Y\leq y_{12},D=1)\\
&&-\mathbb P(Y\leq\min(y_{02},y_{11}))\\
&&-\mathbb P(Y\leq\min(y_{01},y_{12}))\\
&&+\mathbb P(Y\leq y_{01},D=0)+\mathbb P(Y\leq y_{11},D=1)\\
&=&\mathbb P(y_{11}<Y\leq y_{02},D=0)+\mathbb P(y_{01}<Y\leq y_{12},D=1).
\end{eqnarray*}
The latter bounds are not sharp. Indeed:
\begin{eqnarray}
\mathbb P(y_{01}<Y_0\leq y_{02},y_{11}<Y_1\leq y_{12})&=&
\mathbb P(y_{01}<Y_0\leq y_{02},y_{11}<Y_1\leq y_{12},Y_1\leq Y_0)\nonumber\\
&&\hskip25pt+\hskip10pt\mathbb P(y_{01}<Y_0\leq y_{02},y_{11}<Y_1\leq y_{12},Y_1>Y_0)\nonumber\\
&\leq&\mathbb P(\max(y_{01},y_{11})<Y\leq y_{02},D=0) \nonumber\\
&&\hskip25pt+\hskip10pt\mathbb P(\max(y_{01},y_{11})<Y\leq y_{12},D=1),
\nonumber\end{eqnarray}
obtained from Theorem~\ref{thm:func-sharp} (or directly), are sharper unless $y_{01}=y_{11}$.

\subsubsection*{Proof of Theorem~\ref{thm:func-sharp}(1,2)}
The Roy model defined by Assumptions~\ref{ass:roy} and~\ref{ass:sel} can be equivalently recast as a correspondence $G:\mathbb R\times\{0,1\}\rightrightarrows\mathbb R^2$ defined as follows, with the order convention $(Y_0,Y_1)$ for the pair of unobserved variables.
For all $y\in\mathbb R$, 
\begin{eqnarray}
\label{eq:G}
\begin{array}{lll}
G(y,0)&=& \{y\}\times[\underline b,y]\\ \\
G(y,1)&=&[\underline b,y]\times\{y\}.
\end{array}
\end{eqnarray}
Indeed, if $D=0$, by Assumption~\ref{ass:roy}, $Y_0=Y$. By Assumption~\ref{ass:sel}, $Y_1\leq Y$. Hence the set of values compatible with the Roy model specification is $(Y_0,Y_1)\in\{Y\}\times[\underline b,Y]$, as in the definition of $G$. Similarly, if $D=1$, by Assumption~\ref{ass:roy}, $Y_1=Y$. By Assumption~\ref{ass:sel}, $Y_0\leq Y$. Hence the set of values compatible with the Roy model specification is $(Y_0,Y_1)\in[\underline b,Y]\times\{Y\}$.

The collection $(\mu,G,\nu)$, where $\mu$ is the joint distribution of the vector $(Y,D)$ of observable variables and $\nu$ is the joint distribution of the vector $(Y_1,Y_0)$ of unobservable variables, forms a structure in the terminology of \cite{KR:50} extended by \cite{Jovanovic:89}. The correspondence $G$ is non-empty valued and measurable, in the sense that for any open set $\mathcal O\subseteq \mathbb R^2$, $G^{-1}(\mathcal O):=\{(y,d)\in\mathbb R\times\{0,1\} \;\vert\; G(y,d)\cap\mathcal O\ne\varnothing\}$ is a Borel subset of $\mathbb R\times\{0,1\}$. Hence Theorem~1 of \cite{GH:2011} applies and the collection of inequalities
\[
\mu(A)\leq\nu [G^{-1}(A)],\mbox{ for all Borel }A\subseteq\mathbb R^2
\]
define sharp bounds for the joint distribution $\nu$ of the unobservable variables $(Y_1,Y_0)$.

For any Borel $A\subseteq\mathbb R^2,$ 
\begin{eqnarray*}
G^{-1}(A)&=&\{(y,d)\in\mathbb R\times\{0,1\} \;\vert\; G(y,d)\cap A\ne\varnothing\}\\
&=&\{(y,0)\;\vert\; y\in U_{A,0} \}\cup\{(y,1)\;\vert\; y\in U_{A,1}\}.
\end{eqnarray*}
Hence, $\mu(A)\leq\nu [G^{-1}(A)]$ is equivalent to the second inequality in the display of Proposition~\ref{prop:sharp}. The first inequality in that same display is obtained by complementation as follows.
\begin{eqnarray*}
\mu(A^c)\leq\nu[G^{-1}(A^c)]
\hskip15pt \Rightarrow \hskip15pt
\mu (A) &\geq& \nu[\{(y,d)\in\mathbb R\times\{0,1\}\;\vert\; G(y,d)\subseteq A)\}] \\
&=&\nu[\{(y,0)\;\vert\; y\in L_{A,0} \}\cup\{(y,1)\;\vert\; y\in L_{A,1}\}]
\end{eqnarray*}
as required.

\subsubsection*{Proof of Theorem~\ref{thm:func-sharp}(3)}
Let $A$ be an upper set in $\mathbb R^2$. By Assumption~\ref{ass:SMIV}, $\mathbb P((Y_0,Y_1)\in A\vert Z=z)\leq\inf_{\tilde z\geq z}\mathbb P((Y_0,Y_1)\in A\vert Z=\tilde z)$. As shown in the proof of Theorem~\ref{thm:func-sharp}(1,2), $\mathbb P((Y_0,Y_1)\in A\vert Z=\tilde z)\leq \mathbb P(Y\in U_{A,0},D=0\vert Z=\tilde z)+\mathbb P(Y\in U_{A,1},D=1\vert Z=\tilde z)$. Since $A$ is an upper set, $Y\in U_{A,d}$ implies $(Y,Y)\in A$ and the upper bound follows. Similarly, we have $\mathbb P((Y_0,Y_1)\in A\vert Z=z)\geq \mathbb P(Y\in L_{A,0},D=0\vert Z=\tilde z)+\mathbb P(Y\in L_{A,1},D=1\vert Z=\tilde z)$ for any $\tilde z\leq z$. By definition of $L_{A,0}$, $Y\in L_{A,0}$ implies ${Y}\times[\underline b,Y]\subseteq A$. Since $A$ is an upper set, this in turn implies that $[Y,\infty)\times\mathbb R\subseteq A$, hence that $Y\geq\underbar y^A_0$. Reasoning identically for $L_{A,1}$ yields the lower bound and the result follows.

\subsubsection*{Statement of Corollary~\ref{cor:marg}}
\begin{enumerate}
\item Let $(Y_0,Y_1)$ be an arbitrary pair of random variables. Let $Y$ and $D$ satisfy Assumptions~\ref{ass:roy} and~\ref{ass:sel}. Then the distribution functions $F_1$ and $F_0$ of $Y_1$ and $Y_0$ respectively, satisfy
\begin{eqnarray}
\label{eq:fcn}
F_d(y_2)-F_d(y_1)\geq \mathbb P(y_1<Y\leq y_2,D=d)+\mathbb P(Y\leq y_2,D=1-d)1\{y_1\leq\underline b\}
\end{eqnarray}
 for $d=0,1,$ and for all $y_1$ and $y_2$ in $\mathbb R\cup\{\pm\infty\}$, such that $y_1<y_2$.
\item Let $Y$ be an arbitrary random variable and $D$ be a binary random variable. Let $F_1$ and $F_0$ be cumulative distribution functions satisfying (\ref{eq:fcn}). Then there exists a pair $(Y_1,Y_0)$ with cdfs $F_1$ and $F_0$ respectively, such that Assumptions~\ref{ass:roy} and~\ref{ass:sel} are satisfied.
\end{enumerate}

\subsubsection*{Proof of Corollary~\ref{cor:marg}}
\mbox{ }

(1) Validity of the bounds: As shown in the main text, Proposition~\ref{prop:sharp} yields bounds~(\ref{eq:low})-(\ref{eq:up}) and $(\ref{eq:up})$ is redundant. Hence the result.

(2) Sharpness of the bounds:
Let $Y$ and $D$ be given. Let $F_1$ and $F_0$ be cdfs satisfying (\ref{eq:fcn}). We shall construct a pair $(Y_0,Y_1)$ with cdfs $F_0$ and $F_1$ respectively, such that Assumptions~\ref{ass:roy} and~\ref{ass:sel} are satisfied. 

Define $\underbar F_d$ with $y\mapsto \underbar F_d(y)=\mathbb P(Y\leq y,D=d)$ for each $y$. Let $F^{-1}$ be the generalized inverse, defined as $F^{-1}(u)=\inf\{y: F(y)\geq u\}$. Let $U$ be a uniform random variable on $[0,1]$ such that $U< \mathbb P(D=1)\Leftrightarrow D=1$. Define $Y_0$ and $Y_1$ in the following way. When $U<\mathbb P(D=1)$, let $Y_1=\underbar F_1^{-1}(U)$ and $Y_0=(F_0-\underbar F_0)^{-1}(U)$. The latter is well defined, since $F_0\geq\underbar{F}_0$, and $U$ remains in the range of $F_0-\underbar F_0$. Indeed, (\ref{eq:fcn}) implies 
\begin{eqnarray}
\label{eq:red}
F_d(y)\geq\underbar F_d(y)+\underbar F_{1-d}(y),\mbox{ for each }y\in\bar{\mathbb R}.
\end{eqnarray}
Hence, $F_d(y)-\underbar F_d(y)\geq \mathbb P(Y\leq y,D=1-d)$, hence, in particular, $(F_0-\underbar F_0)(+\infty)\geq\mathbb P(D=1)$. For $U>\mathbb P(D=1)$, let  $Y_0=\underbar F_0^{-1}(U-\mathbb P(D=1))$ and $Y_1=(F_1-\underbar F_1)^{-1}(U-\mathbb P(D=1))$. The latter is well defined because, as before, $(F_1-\underbar F_1)(+\infty)\geq\mathbb P(D=0)$.

We first verify Assumption~\ref{ass:sel}. Note first that Assumption~\ref{ass:sel} is equivalent to $D=d\Rightarrow Y_d\geq Y_{1-d}$ for $d=0,1$. Hence, we need to show that $U<\mathbb P(D=1)\Rightarrow Y_1\geq Y_0$ and $U>\mathbb P(D=1)\Rightarrow Y_1\leq Y_0$. By symmetry, we only show the first implication. Suppose $U<\mathbb P(D=1)$. If $U$ is a continuity value of $\underbar F_1$, then $U=\underbar F_1(Y_1)$. By (\ref{eq:red}), $\underbar F_1\leq F_0-\underbar F_0$. Hence, $U=\underbar F_1(Y_1)\leq(F_0-\underbar F_0)(Y_1)$. So if we can show right-continuity and monotonicity of $F_d-\underbar F_d$, then $Y_0=(F_0-\underbar F_0)^{-1}(U)\leq Y_1$ as required.
Now, monotonicity of $F_d-\underbar F_d$ follows immediately from (\ref{eq:fcn}) and right continuity of $F_d-\underbar F_d$ from that of $F_d$ and $\underbar F_d$. If the distribution of $Y_1$ has an atom at $\underbar F_1^{-1}(U)$, then, by right-continuity of $\underbar F_1$, $U\leq\underbar F_1(Y_1)\leq (F_0-\underbar F_0)(Y_1)$, so that, by right continuity and monotonicity of $F_0-\underbar F_0$, we have $Y_0=(F_0-\underbar F_0)^{-1}(U)\leq Y_1$ as required.

We now verify Assumption~\ref{ass:roy}. We need to show that for each $d=1,0$, $\mathbb P(Y_d\leq y,D=d)=\underbar F_d(y)$. By symmetry, we only deal with $Y_1$. By monotonicity and right continuity of $\underbar F_1$, $\underbar F_1^{-1}(U)\leq y\Leftrightarrow U\leq \underbar F_1(y)$ (Proposition 1(5) in \cite{EH:2013}). Hence, we have the following as required.
\[
\mathbb P(Y_1\leq y,D=1)=\mathbb P(\underbar F_1^{-1}(U)\leq y,U<\mathbb P(D=1))=\mathbb P(U\leq\underbar F_1(y),U<\mathbb P(D=1))=\underbar F_1(y).
\] 
Finally, we need to verify that $Y_1$ and $Y_0$ do indeed have the announced distributions. We shall show that the cdf of $Y_1$ is indeed $F_1$. Reasoning as above, we have the following.
\begin{eqnarray*}
\mathbb P(Y_1\leq y,D=0)&=&\mathbb P((F_1-\underbar F_1)^{-1}(U-\mathbb P(D=1))\leq y,U>\mathbb P(D=1))\\
&=&\mathbb P(U\leq (F_1-\underbar F_1)(y)+\mathbb P(D=1),U<\mathbb P(D=1))\\
&=&(F_1-\underbar F_1)(y).
\end{eqnarray*}
Therefore $\mathbb  P(Y_1\leq y)=\mathbb P(Y_1\leq y,D=1)+\mathbb P(Y_1\leq y, D=0)=\underbar F_1(y)+(F_1-\underbar F_1)(y)=F_1(y)$ as required.

\subsection{Functional features of potential outcomes}

\subsubsection*{Proof of Proposition~\ref{prop:ineq}}
Let $0<q_1<q_2<1$ and let $y_1$ and $y_2$ be the $q_1$ and $q_2$ quantiles of the distribution of outcomes in Sector~$d$ for the self-selected economy. The following holds.
\begin{eqnarray*}
\mathbb P(Y\leq y_2\vert D=d)-\mathbb P(Y\leq y_1\vert D=d)=
\frac{1}{\mathbb P(D=d)}\left(
\underbar F_d(y_2)-\underbar F_d(y_1)
\right)
\geq \underbar F_d(y_2)-\underbar F_d(y_1).
\end{eqnarray*}
Hence, $q_2-q_1\geq \underbar F_d(y_2)-\underbar F_d(y_1)$.
In addition, for any $y\in\mathbb R$, 
\begin{eqnarray*}
\mathbb P(Y\leq y\vert D=d)=\frac{\underbar F_d(y)}{\mathbb P(D=d)}=\underbar F_d(y)+P(D=1-d)\frac{\underbar F_d(y)}{\mathbb P(D=d)}\leq \bar F_d(y).
\end{eqnarray*}
Finally, under the stochastic dominance condition, $F(y_j)=\mathbb P(Y\leq y_j)\leq\mathbb P(Y\leq y_j\vert D=d)$ for $j=0,1$. 
Since the sharp upper bound for the $(q_1,q_2)$-interquantile range of the distribution of $Y_d$ is given by
\begin{eqnarray*}
\bar{\mbox{IQR}}(q_1,q_2,F_d)=\max\left\{y_2-y_1\left| 
\begin{array}{l}
F(y_1)\leq q_1\leq \bar F_d(y_1),\\
F(y_2)\leq q_2\leq\bar F_d(y_2),\\
q_2-q_1\geq \underbar F_d(y_2)-\underbar F_d(y_1),
\end{array}
\right.\right\}
\end{eqnarray*}
these three inequalities imply that the interquantile range $y_2-y_1$ satisfies the sharp bounds on the interquantile range for the distribution of $Y_d$.

However, if we relax the first order stochastic dominance condition, we now show that there exist situations, where the interquantile range in Sector $d$ under self-selection is strictly larger than the upper bound for the corresponding interquantile range of the distribution of potential outcomes under random assignment. Let $\mathbb P(Y\leq y)$ be continuous. Let $y_{02}$ be defined by $\mathbb P(Y\leq y_{02},D=d)+\mathbb P(D=1-d)=q_2$. Finally, suppose that $D=1-d\Rightarrow Y\leq y_{02}$, so that $\mathbb P(Y\leq y)$ coincides with $\mathbb P(Y\leq y,D=d)+\mathbb P(D=1-d)$ on the right of $y_{02}$. Then the upper bound for the interquantile range of $Y_d$ is $\underbar F_d(q_2-\mathbb P(D=d))-\underbar F_d(q_1-\mathbb P(D=d))$, which can be made lower than the interquantile range for $Y\vert D=d$, namely $\mathbb P(D=d)\left(\underbar F_d(q_2)-\underbar F_d(q_1)\right)$, with a suitable choice of slope for $\mathbb P(Y\leq y,D=d)$. 
}

\end{appendix}



\bibliography{Roy}
\bibliographystyle{plain}

\newpage


{\sc
\begin{table}
\begin{center}
\begin{tabular}{llrrr}
\multicolumn{5}{l}{CANADA}\\
\hline\\
& & {\sc stem} & other & total \\ \\
\hline\\ \\
women & & & & \\
& minority &151 &841 & 992\\ \\
& white & 298& 4,328& 4,626\\ \\
& total &449 &5,169 & 5,618\\
men & & & & \\
& minority & 445&528 & 983\\ \\
& white & 1,052& 2,265& 3,317\\ \\
& total & 1,507&2,793 &4,300 \\ \\ \\
total & &1,956& 7,962& 9,918 \\ \\
\hline \\ \\ \\ \\ \\ \\ 
\multicolumn{5}{l}{GERMANY} \\
\hline\\
& & stem & other & total \\ \\
\hline\\ \\
women & & & & \\
& minority &243 & 263& 506\\ \\
& white &1,398 & 2,542& 3,940\\ \\
& total & 1,641&2,805 & 4,446\\
men & & & & \\
& minority &276 &74 &350 \\ \\
& white & 2,170& 751& 2,921\\ \\
& total & 2,446&825 &3,271 \\\\\\
total & &4,087 &3,630 &7,717 \\ \\
\hline
\end{tabular}
\end{center}
\vskip25pt
\caption{Major Choice}
\label{table:STEM}
\end{table}
}


{\sc
\begin{table}
\vskip50pt
\begin{center}
\begin{tabular}{llrrrrrrr}
\multicolumn{5}{l}{CANADA}\\
\hline\\
& & permanent & other & total \\ \\
\hline\\ \\
women & & & & \\
& stem & 369& 79& 448 \\ \\
& other &4,009 &995 &5,004  \\ \\
& total & 4,378& 1,074& 5,452 \\
men & & & &  \\
& stem &1,332 &135 &1,467  \\ \\
& other & 2,168& 448&2,616  \\ \\
& total & 3,500&583 &4,083  \\ \\
total & & 7,878& 1,657& 9,535  \\ \\
\hline \\ \\ \\ \\ \\ \\ 
\multicolumn{5}{l}{GERMANY} \\
\hline\\
& & permanent & other & total \\ \\
\hline\\ \\
women & & & &\\
& stem & 443 &804&1,247\\ \\
& other & 499  &1702 &2,201\\ \\
& total & 942 &2,506 &3,448\\
men & & & &\\
& stem & 769 &1,100 &1,869\\ \\
& other & 153 &508 &661\\ \\
& total & 922 &1,608 &2,530\\ \\
total & & 1,864 &4,114 &5,978\\ \\
\hline
\end{tabular}
\end{center}
\vskip25pt
\caption{Permanent employment 1 year (Germany) and 3 years (Canada) after graduation. Each entry is the number of individuals in that category.}
\label{table:Perm}
\end{table}
}


{\sc
\begin{table}
\vskip80pt
\begin{center}
\begin{tabular}{llcc}\multicolumn{4}{r}{}\\\\
\multicolumn{4}{l}{CANADA}\\
\hline\\
& & \multicolumn{2}{c}{}  \\ \\
\hline \\ \\
& & Qu\'ebec & Rest \\ \\
women & & &    \\
& minority &1.784 (.507) &1.680 (.552) \\ \\
& white & 1.591 (.611)& 1.593 (.601) \\
men & & &  \\
& minority & 1.691 (.538)& 1.629 (.575) \\ \\
& white  & 1.635 (.560)& 1.574 (.618) \\ \\
\hline \\ \\ \\ \\ \\ \\ 
\multicolumn{4}{l}{GERMANY} \\
\hline\\
& & \multicolumn{2}{c}{}  \\ \\
\hline \\ \\
& & east & west   \\ \\
women & & & \\
& minority & - & 2.102 (.054)\\ \\
& white & 1.842 (.036) \hskip50pt & 1.793 (.024) \\
men & & &  \\
& minority & - & 2.165 (.066)\\ \\
& white & 2.133 (.051) & 2.090 (.027) \\ \\
\hline
\end{tabular}
\end{center}
\vskip25pt
\caption{ {\scriptsize Mean survey responses (with standard deviations in parentheses) to a question on the importance of labor market considerations on major choice, from $0$ for ``not important'' to $3$ for ``very important'' in Canada, and from $0$ for ``not at all important,'' to $4$ for ``very important'' for Germany.}}
\label{table:LM}
\end{table}
}


{\sc
\begin{table}
\vskip80pt
\begin{center}
\begin{tabular}{llcccccc}
\multicolumn{8}{r}{SMIV: Mother's education}\\\\\\\\
\multicolumn{8}{l}{CANADA}\\
\hline\\
& & \multicolumn{6}{c}{target variable}  \\ \\
& & \multicolumn{2}{c}{income} & \multicolumn{2}{c}{permanent} & \multicolumn{2}{c}{related} \\ \\
\hline \\ \\
& & Qu\'ebec & Rest of& Qu\'ebec & Rest of& Qu\'ebec & Rest of\\
& & & Canada &  & Canada &  & Canada \\ \\
women & & & & & & &  \\
& minority & & & & & &\\ \\
& white & & R$^{**}$&  &R$^{**}$ &R$^{**}$ & \\
men & & & &  & & & \\
& minority & & &  &  & &\\ \\
& white  & &  & & & &\\ \\
\hline \\ \\ \\ \\ \\ \\ 
\multicolumn{8}{l}{GERMANY} \\
\hline\\
& & \multicolumn{6}{c}{target variable}  \\ \\
& & \multicolumn{2}{c}{income} & \multicolumn{2}{c}{permanent} & \multicolumn{2}{c}{related} \\ \\
\hline \\ \\
& & east & west & east & west & east & west \\ \\
women & & & & & &  &\\
& minority & - & &- & & -& \\ \\
& white & & R$^{***}$& R$^{*}$ &R$^{***}$ & & \\
men & & & &  & & & \\
& minority &- & & - & & -&\\ \\
& white & & R$^{***}$&  &R$^{***}$& & \\ \\
\hline
\end{tabular}
\end{center}
\vskip25pt
\caption{Test of Roy with imperfect foresight with mother's education as an instrument. ``-'' indicates the test was not applied to that category because of low sample size.}
\label{table:IFm}
\end{table}
}


{\sc
\begin{table}
\vskip80pt
\begin{center}
\begin{tabular}{llcccccc}
\multicolumn{8}{r}{SMIV: Mother's education}\\\\\\\\
\multicolumn{8}{l}{CANADA}\\
\hline\\
& & \multicolumn{6}{c}{target variable}  \\ \\
& & \multicolumn{2}{c}{income} & \multicolumn{2}{c}{lexicographic} & \multicolumn{2}{c}{related}  \\ \\
\hline \\ \\
& & Qu\'ebec & Rest of& Qu\'ebec & Rest of& Qu\'ebec & Rest of\\ 
& &  & Canada & & Canada  & & Canada \\ \\
women & & & & & &&  \\
& minority & & & & &&\\ \\
& white & & R$^{**}$&  R$^{*}$ & R$^{**}$ &R$^{*}$&\\
men & & & &  & &&\\
& minority & &  & &&&\\ \\
& white  & &  & &&&\\ \\
\hline \\ \\ \\ \\ \\ \\ 
\multicolumn{8}{l}{GERMANY} \\
\hline\\
& & \multicolumn{6}{c}{target variable}  \\ \\
& & \multicolumn{2}{c}{income} & \multicolumn{2}{c}{lexicographic} & \multicolumn{2}{c}{related} \\ \\
\hline \\ \\
& & east & west & east & west  & east & west \\ \\
women & & & &  &&&\\
& minority &-  & & -& &-&\\ \\
& white &R$^{*}$ & R$^{***}$ & R$^{*}$& R$^{***}$ &&\\
men & & & & & &&\\
& minority &  - &  & -& &-&\\ \\
& white & & R$^{***}$ & & R$^{***}$ &&\\ \\
\hline
\end{tabular}
\end{center}
\vskip25pt
\caption{Test of Roy with perfect foresight with mother's education as an instrument. {\sc lexicographic} refers to the vector of variables {\sc (permanent, related)} ordered lexicographically. ``-'' indicates the test was not applied to that category because of low sample size.}
\label{table:Rm}
\end{table}
}


{\sc
\begin{table}
\vskip80pt
\begin{center}
\begin{tabular}{llcccccc}
\multicolumn{8}{r}{SMIV: Mother's education}\\
\multicolumn{8}{r}{and feminization of STEM}\\\\\\\\
\multicolumn{8}{l}{CANADA}\\
\hline\\
& & \multicolumn{6}{c}{target variable}  \\ \\
& & \multicolumn{2}{c}{income} & \multicolumn{2}{c}{permanent} & \multicolumn{2}{c}{related} \\ \\
\hline \\ \\
& & Qu\'ebec & Rest of& Qu\'ebec & Rest of& Qu\'ebec & Rest of\\
& & & Canada &  & Canada &  & Canada \\ \\
women & & & & & & &  \\
& minority & & & & R$^{**}$ & &\\ \\
& white & & & & R$^{**}$& R$^{***}$& \\ \\ \\ \\
\hline \\ \\ \\ \\ \\ \\ 
\\ \\ \\ \\ \\ \\ \\ \\
\multicolumn{8}{l}{GERMANY} \\
\hline\\
& & \multicolumn{6}{c}{target variable}  \\ \\
& & \multicolumn{2}{c}{income} & \multicolumn{2}{c}{permanent} & \multicolumn{2}{c}{related} \\ \\
\hline \\ \\
& & east & west & east & west & east & west \\ \\
women & & & & & &  &\\
& minority & -& &- & & -& \\ \\
& white & R$^{***}$& R$^{***}$& R$^{***}$ &R$^{***}$ & & \\
\\ \\ \\
\hline
\end{tabular}
\end{center}
\vskip25pt
\caption{Test of Roy with imperfect foresight with mother's education and feminization of STEM faculty as instruments. ``-'' indicates the test was not applied to that category because of low sample size.}
\label{table:IFmf}
\end{table}
}


{\sc\scriptsize
\begin{table}
\vskip80pt
\begin{center}
\begin{tabular}{llcccccc}
\multicolumn{8}{r}{SMIV: Mother's education}\\
\multicolumn{8}{r}{and feminization of STEM}\\\\\\\\
\multicolumn{8}{l}{CANADA}\\
\hline\\
& & \multicolumn{6}{c}{target variable}  \\ \\
& & \multicolumn{2}{c}{income} & \multicolumn{2}{c}{lexicographic} & \multicolumn{2}{c}{related} \\ \\
\hline \\ \\
& & Qu\'ebec & Rest of& Qu\'ebec & Rest of& Qu\'ebec & Rest of\\ 
& &  & Canada & & Canada & & Canada \\ \\
women & & & & &  && \\
& minority & & & & R$^{**}$ &&\\ \\
& white & & & R$^{**}$  & R$^{**}$&R$^{***}$&\\ \\ \\ \\ \\
\hline \\ \\ \\ \\ \\ \\ 
\\ \\ \\ \\ \\ \\ \\ \\
\multicolumn{8}{l}{GERMANY} \\
\hline\\
& & \multicolumn{6}{c}{target variable}  \\ \\
& & \multicolumn{2}{c}{income} & \multicolumn{2}{c}{lexicographic} & \multicolumn{2}{c}{related} \\ \\
\hline \\ \\
& & east & west & east & west  & east & west  \\ \\
women & & & &  &&&\\
& minority &  -& & -& &-& \\ \\
& white &R$^{***}$ & R$^{***}$ & R$^{**}$&R$^{***}$ &&\\ \\ \\ \\ \\
\hline
\end{tabular}
\end{center}
\vskip25pt
\caption{Test of Roy with perfect foresight with mother's education and feminization of stem as instruments. {\sc lexicographic} refers to the vector of variables {\sc (permanent, related)} ordered lexicographically. ``-'' indicates the test was not applied to that category because of low sample size.}
\label{table:Rmf}
\end{table}
}


\begin{figure}[htbp]
\centering
\subfloat[Germany]{
  \includegraphics[width=0.8\textwidth]{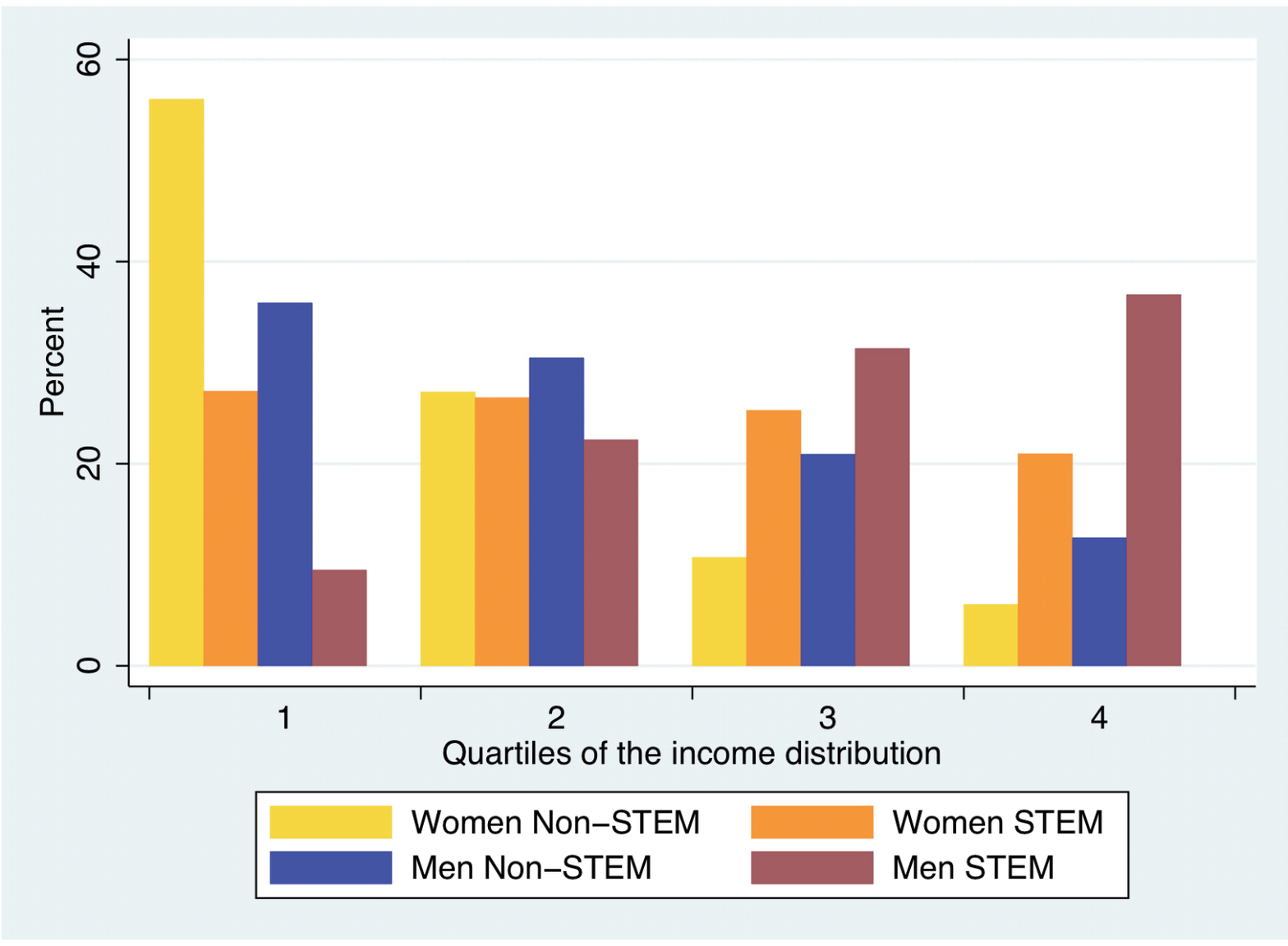}
}

\subfloat[Canada]{
   \includegraphics[width=0.8\textwidth]{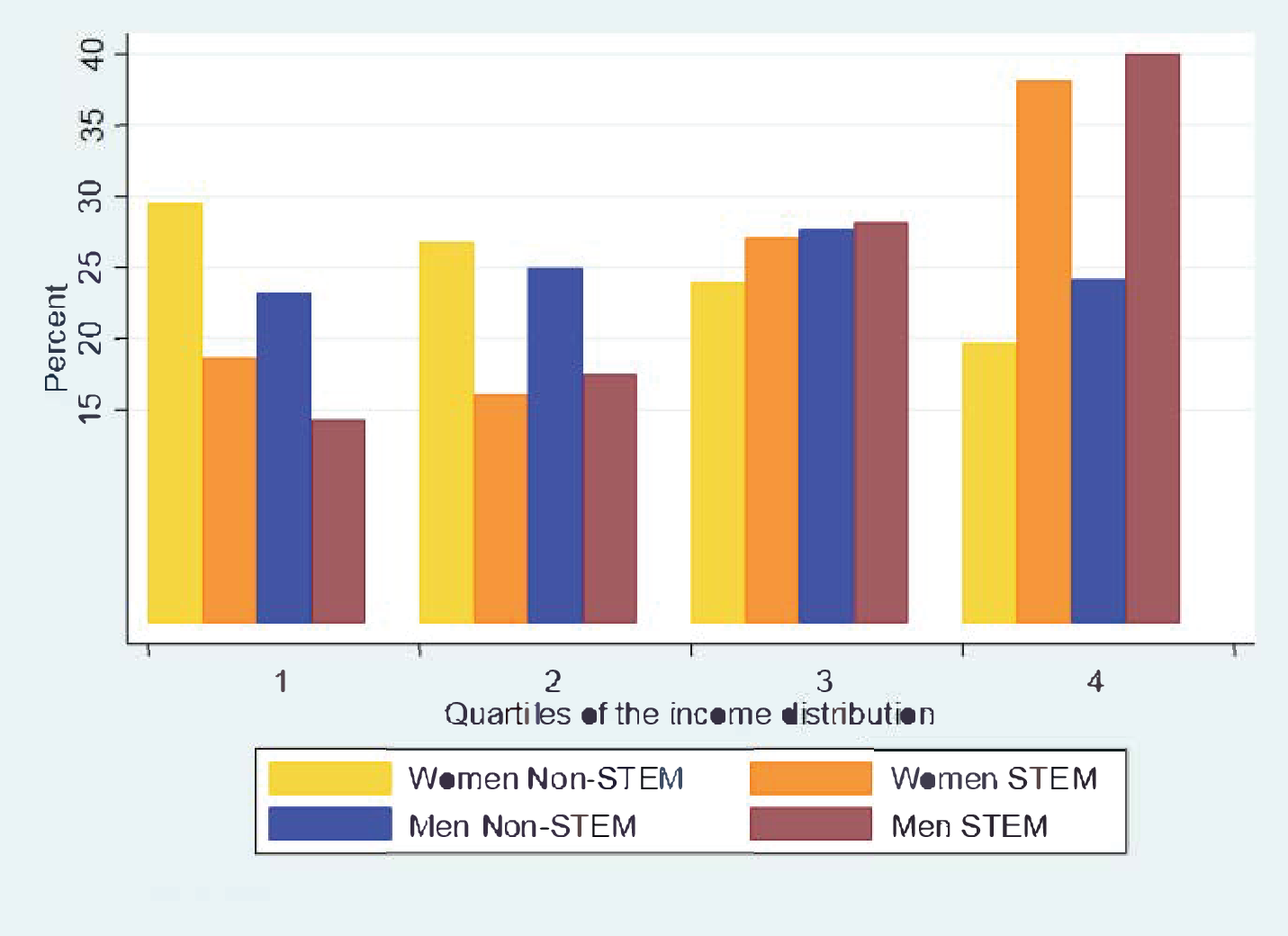}
}
\caption{Quartiles of income distributions by gender and major choice.}
\label{fig:Quartiles}
\end{figure}


\begin{figure}[htbp]
\centering
\vskip25pt
\subfloat[Mother's education on women.]{
   \includegraphics[width=0.75\textwidth]{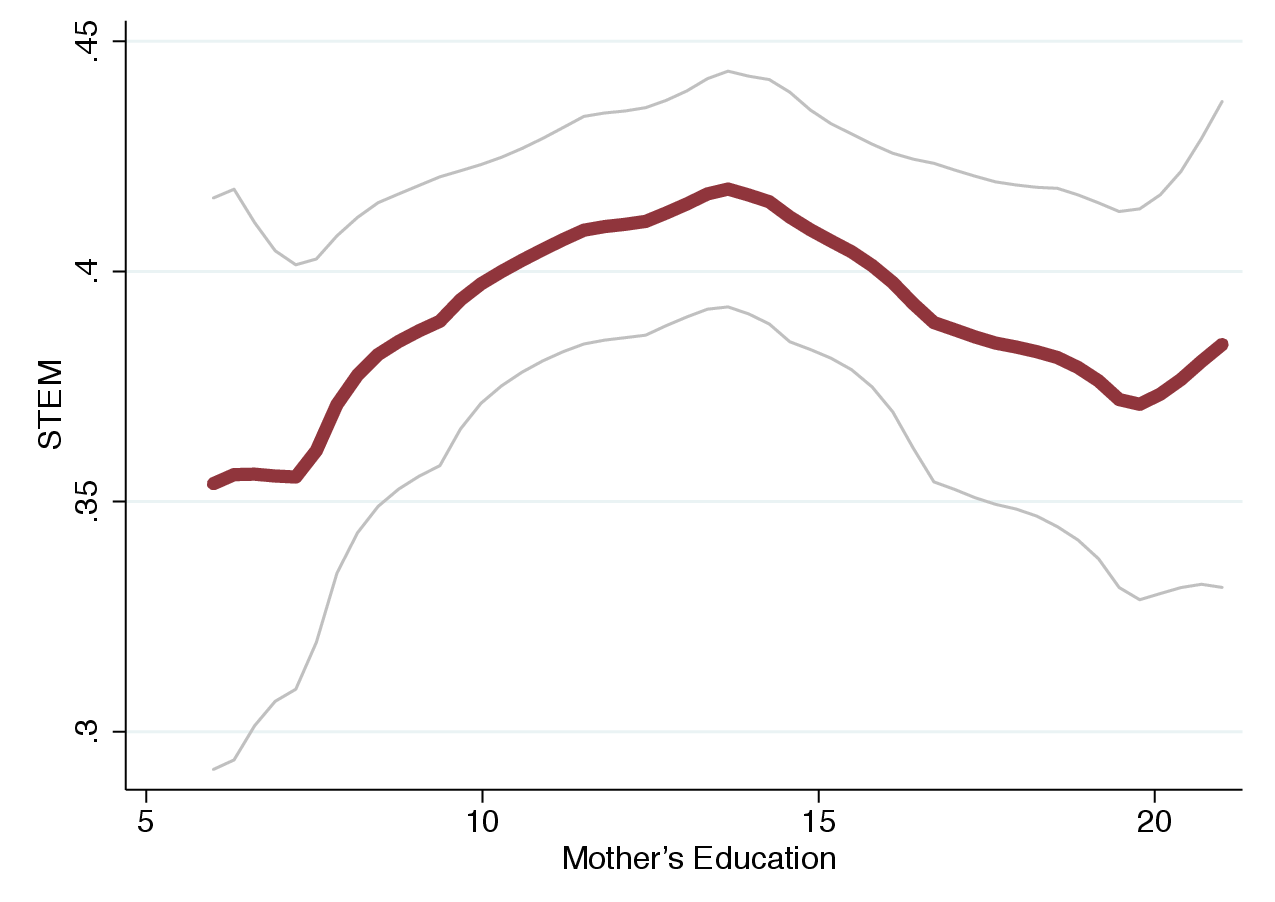}
}
\vskip50pt
\subfloat[Feminization of STEM on women]{
   \includegraphics[width=0.75\textwidth]{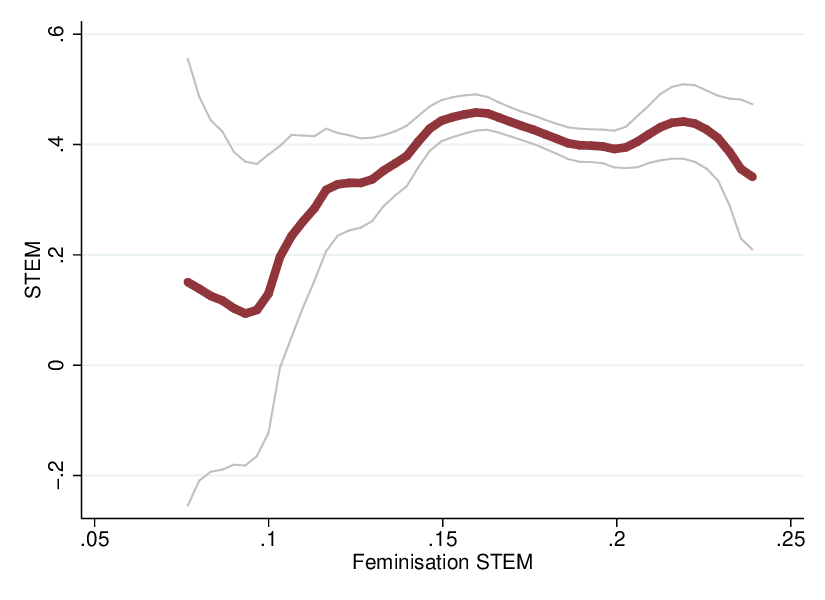}
}
\vskip25pt
\caption{{\scriptsize Effect of instruments on STEM choices by white residents of the former FRG.  The brown line is the point estimator and the grey lines are the~$95\%$ confidence bands.}}
\label{fig:INS}
\end{figure}



\begin{figure}[htbp]
\centering
\subfloat[White Men in West Germany]{
 \includegraphics[width=0.8\textwidth]{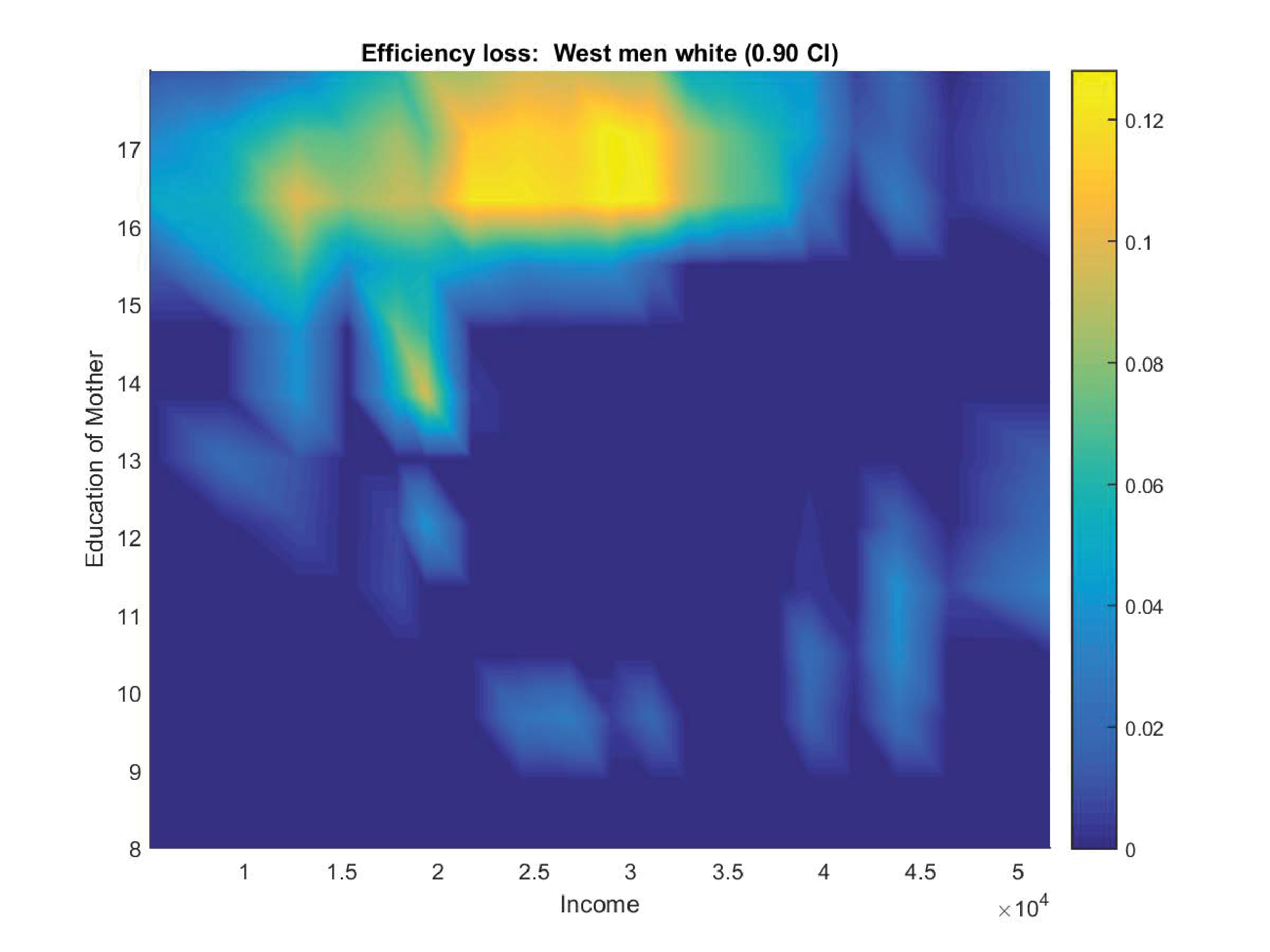}
}
 
\subfloat[White Women in West Germany]{
  \includegraphics[width=0.8\textwidth]{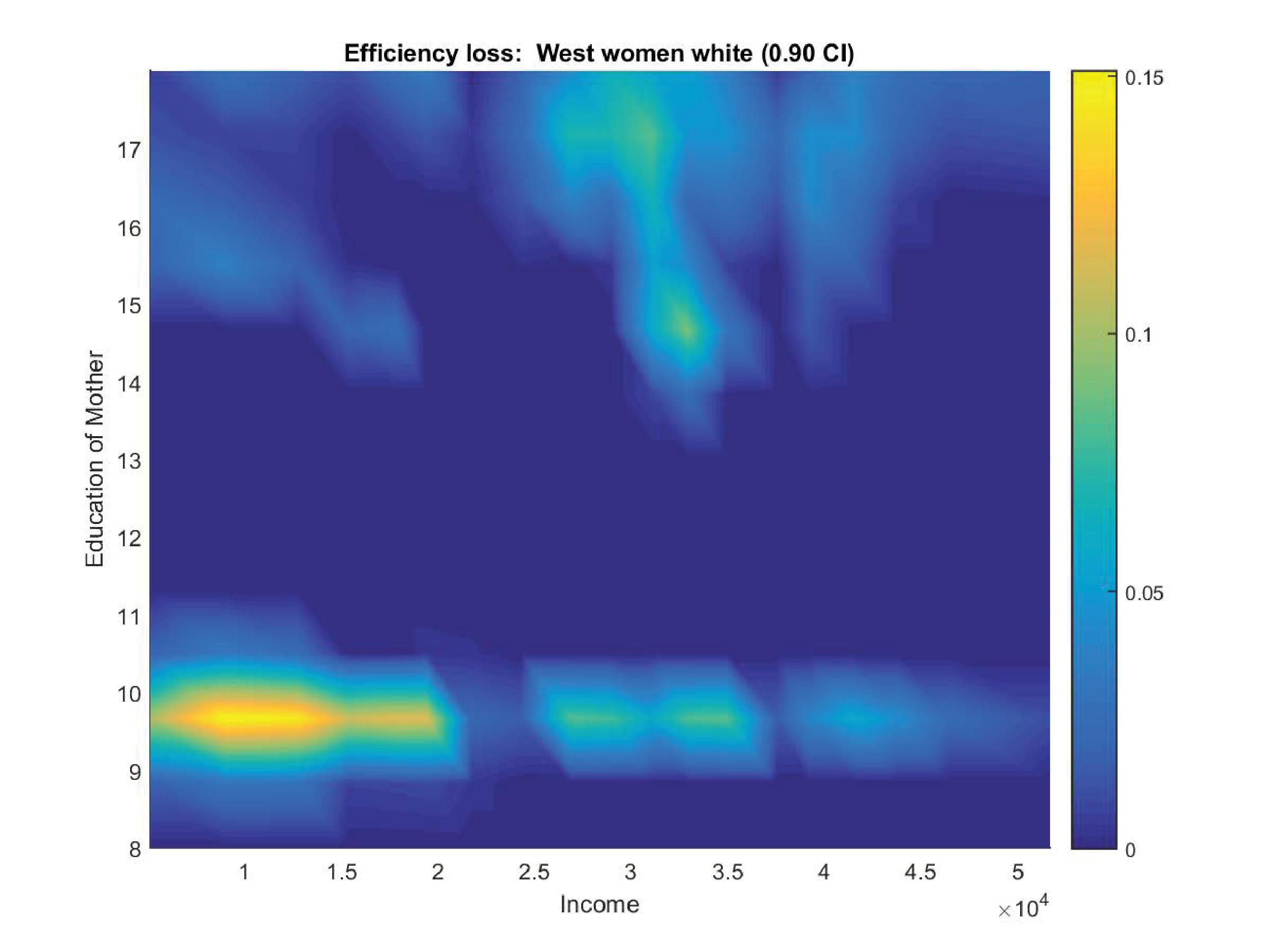}
}
\caption{Lower bound of the 90\% confidence interval for efficiency loss for white men and women from the former FRG using mother's education as a SMIV instrument.}
\label{fig:EL-motherW}
\end{figure}


\begin{figure}[htbp]
\centering
\subfloat[Germany]{
  \includegraphics[width=0.8\textwidth]{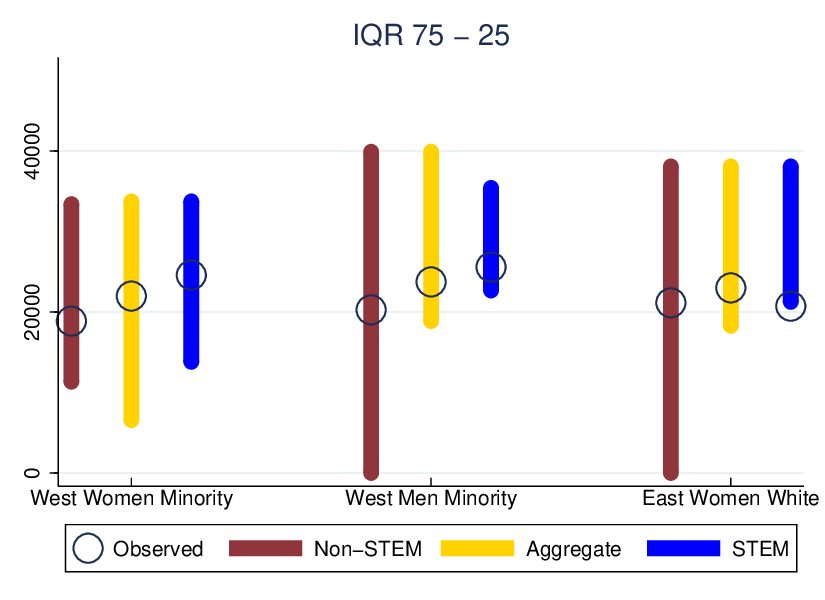}
}

\subfloat[Canada]{
   \includegraphics[width=0.8\textwidth]{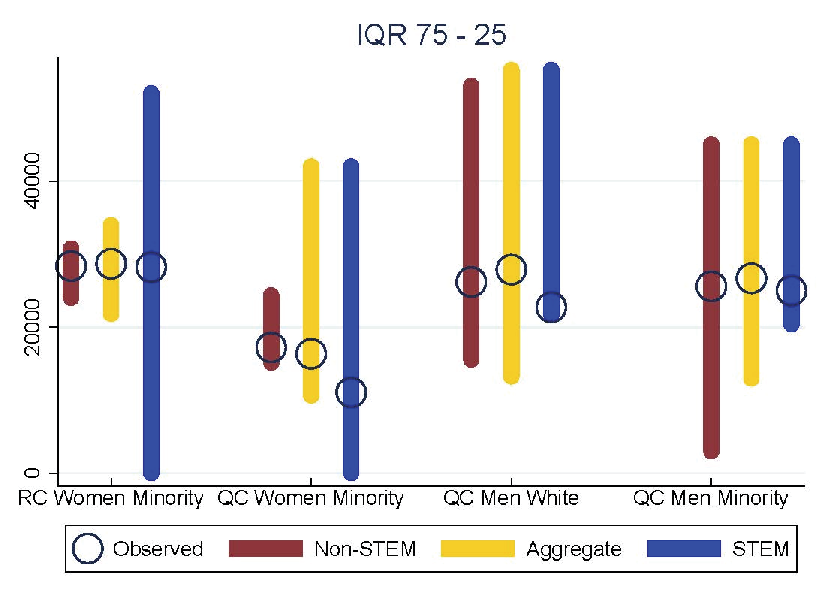}
}
\caption{Confidence bands for the interquartile range under the Roy model assumption for categories of individuals, where the latter is not rejected. ``QC'' and``RC'' stand for ``Qu\'ebec'' and ``Rest of Canada,'' respectively.}
\label{fig:IQR}
\end{figure}

\end{document}